\documentclass[twocolumn]{aa}
\usepackage{graphicx}
\usepackage{float}
\usepackage{caption}
\usepackage{txfonts}
\usepackage{booktabs}
\usepackage{color}
\usepackage[dvipsnames]{xcolor}
\usepackage{natbib,twoopt}
\usepackage{lscape}
\usepackage{caption}
\usepackage{empheq}
\usepackage{subcaption}
\usepackage[symbol]{footmisc}
\usepackage{tablefootnote}
\usepackage[normalem]{ulem}
\usepackage[bookmarks=true, pdfnewwindow=true, colorlinks=true, allcolors=blue, breaklinks=true, final=true, citecolor=blue]{hyperref}
\usepackage{orcidlink}
\bibpunct{(}{)}{;}{a}{}{,} 
\newcommand\degree{\mbox{$^\circ$}}%
\newcommand\farcd{\mbox{$.\!\!^{\circ}$}}%
\newcommand{\gpsc}{\mbox{g~cm$^{-2}$}}%
\newcommand{\scpg}{\mbox{cm$^{2}$~g$^{-1}$}}
\newcommand{\kms}{\mbox{km\,s$^{-1}$}}%
\newcommand{\msun}{\mbox{$M_\odot$}}%
\newcommand{\lmsun}{\mbox{$L_\odot/M_\odot$}}%
\newcommand{\jybeam}{\mbox{Jy\,beam$^{-1}$}}%
\newcommand{\mjybeam}{\mbox{mJy\,beam$^{-1}$}}%
\newcommand{\mjysr}{\mbox{MJy~sr$^{-1}$}}

\newcommand{\um}{\mbox{$\mu$m}}%

\newif\ifmarkrevisions
\markrevisionsfalse
\DeclareRobustCommand{\revise}[1]{\ifmarkrevisions\ifmmode{\color{red}#1}\else{\color{red}\bfseries #1}\fi\else #1\fi}
\addto\captionsenglish{%
}

\begin{document}

\AddToHookNext{shipout/foreground}{%
  \put(0,0){\kern\dimexpr\paperwidth-37mm\relax%
    \raisebox{\dimexpr-25mm-\height\relax}{\href{https://xfengwei.github.io/LANCET/}{%
      \includegraphics[width=23mm]{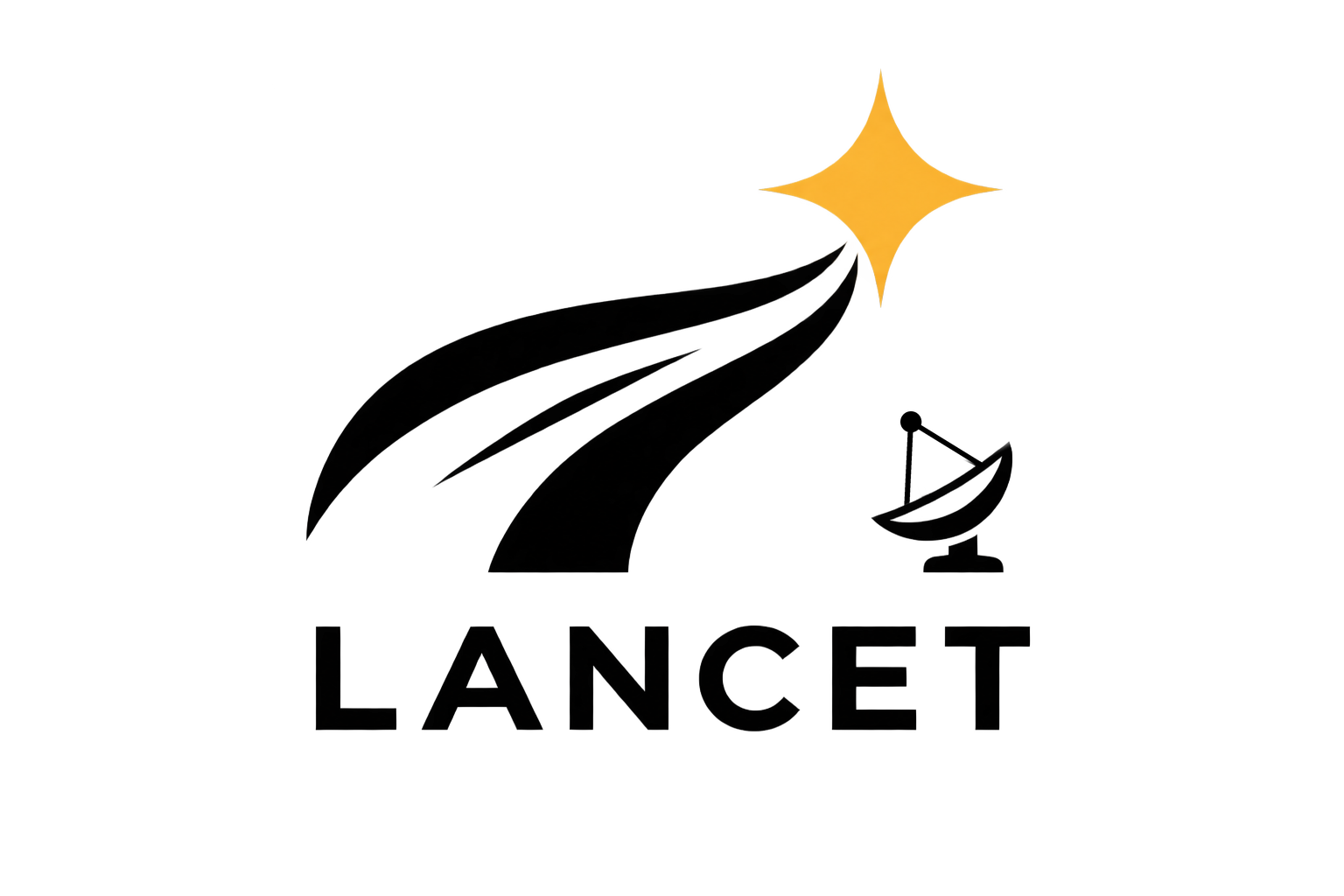}}}}%
}

\title{Linear filament and nested cluster evolution tomography (LANCET)}

\subtitle{II. Hierarchical mass assembly from filamentary clouds to dense cores in G316.8
}

%

\author{Fengwei Xu \inst{\ref{mpia}} \thanks{fengweilookuper@gmail.com, fengwei@mpia.de (PFE fellow)} \orcidlink{0000-0001-5950-1932} \and
Henrik Beuther \inst{\ref{mpia}} \and
Shuting Lin \inst{\ref{xmu}} \and
Hauyu Baobab Liu \inst{\ref{nsysu}}\fnmsep\inst{\ref{cag}} \and
Ke Wang \inst{\ref{kiaa}} \and
Sihan Jiao \inst{\ref{naoc}}\fnmsep\inst{\ref{mpia}} \and
Hao He \inst{\ref{ubonn}} \and
Nicola Schneider \inst{\ref{uzk}} \orcidlink{0000-0003-3485-6678} \and
Adam Ginsburg \inst{\ref{uf}} \orcidlink{0000-0001-6431-9633} \and
Qizhou Zhang \inst{\ref{cfa}} \and
Th. Henning \inst{\ref{mpia}} \and
Guido Garay \inst{\ref{dauc}}\fnmsep\inst{\ref{cassaca}} \orcidlink{0000-0003-1649-7958} \and
Floris F.~S. van der Tak \inst{\ref{sron}} \and
C.-C. Chang \inst{\ref{nsysu}} \and
Zhi-Yu Zhang \inst{\ref{nju}}\fnmsep\inst{\ref{klmaa}} \and
Zipeng Hu \inst{\ref{kiaa}} \orcidlink{0000-0002-3758-552X} \and
Hendrik Linz \inst{\ref{mpia}} \and
F. Wyrowski \inst{\ref{mpifr}} \and
Mark J. McCaughrean \inst{\ref{mpia}} \orcidlink{0000-0002-1452-5268} \and
Keyun Su \inst{\ref{kiaa}} \and
Chang-Xiang Xu \inst{\ref{pku}}
}

\institute{
\label{mpia} Max Planck Institute for Astronomy, K{\"o}nigstuhl 17, 69117 Heidelberg, Germany \and
\label{xmu} Department of Astronomy, Xiamen University, Zengcuo'an West Road, Xiamen, 361005, PR China \and 
\label{nsysu} Department of Physics, National Sun Yat-Sen University, No. 70, Lien-Hai Road, Kaohsiung City 80424, Taiwan, R.O.C. \and 
\label{cag} Center of Astronomy and Gravitation, National Taiwan Normal University, Taipei 116, Taiwan \and 
\label{kiaa} Kavli Institute for Astronomy and Astrophysics, Peking University, Beijing 100871, PR China \and
\label{naoc} National Astronomical Observatories, Chinese Academy of Sciences, Beijing 100101, PR China \and
\label{ubonn} Argelander-Institut f{\"u}r Astronomie, Universit{\"a}t Bonn, Auf dem H{\"u}gel 71, 53121 Bonn, Germany \and
\label{uzk} I. Physikalisches Institut, Universit{\"a}t zu K{\"o}ln, Z{\"u}lpicher Stra{\ss}e 77, 50937 Cologne, Germany \and
\label{uf} Department of Astronomy, University of Florida, 211 Bryant Space Science Center, P.O. Box 112055, Gainesville, FL 32611-2055, USA \and
\label{cfa} Center for Astrophysics $|$ Harvard \& Smithsonian, 60 Garden Street, Cambridge, MA 02138, USA \and
\label{dauc} Departamento de Astronom{\'i}a, Universidad de Chile, Las Condes, 7591245 Santiago, Chile \and
\label{cassaca} Chinese Academy of Sciences South America Center for Astronomy, National Astronomical Observatories, Chinese Academy of Sciences, Beijing, 100101, PR China \and
\label{sron} SRON Netherlands Institute for Space Research \& Kapteyn Astronomical Institute, University of Groningen, 9747 AD Groningen, The Netherlands \and
\label{nju} School of Astronomy and Space Science, Nanjing University, Nanjing, Jiangsu 210093, PR China \and
\label{klmaa} Key Laboratory of Modern Astronomy and Astrophysics (Nanjing University), Ministry of Education, Nanjing 210093, PR China \and
\label{mpifr} Max-Planck-Institut f{\"u}r Radioastronomie, Auf dem H{\"u}gel 69, D-53121 Bonn, Germany \and
\label{pku} School of Physics, Peking University, Beijing 100871, PR China
}

\date{Manuscript is updated to \today}

\titlerunning{LANCET. II. - Hierarchical mass assembly in G316.8}

\authorrunning{Xu et al.}

%

\abstract
{
Star cluster formation involves the hierarchical redistribution of gas from giant molecular clouds through dense clumps to individual cores, yet observational constraints simultaneously across these scales remain limited. As part of the Linear Filament and Nested Cluster Evolution Tomography (LANCET) project, we present a multiscale 1.3 mm continuum image of the massive Galactic molecular filament G316.8 by combining Atacama Large Millimeter/submillimeter Array (ALMA) interferometric data with model-reconstructed total-power emission. The resulting map traces extended emission across the 14-pc filament while resolving thousand-au dense cores. Its coverage achieves an in situ evolutionary sequence from a northern infrared-dark region, through an intermediate high-mass star-forming region, to a southern H{\sc ii} region. Using the \texttt{getsf} algorithm, we identify 231 dense cores, separating them from their surrounding intercore medium as well as larger-scale cloud material. The total-power contribution supplies extended emission for the cloud-scale context and the foreground/background used in the core optical-depth calculation. The mass transfer efficiency from cloud ($\sim4$~pc) to core ($\sim0.01$~pc) scales increases from $0.4\%$ to $3.1\%$ along the evolutionary sequence. Including fragments ($\sim0.15$~pc) identified in ACA 7~m-only data, we find that the cloud-to-fragment efficiency rises from $0.7\%$ to $16.5\%$, whereas the fragment-to-core efficiency decreases from \revise{$55.6\%$} to $18.7\%$ from the young to the evolved region. These efficiencies are ratios of the observed masses at the corresponding scales, tracing how mass is distributed across the hierarchy. We characterize each core's environment using median values within a 0.4 pc diameter aperture. The intercore surface density sampled at each core position on the $\sim0.009$~pc beam scale correlates with the median of the same map within the 0.4~pc aperture (coefficient \revise{0.70}; power-law index \revise{$1.24\pm0.07$}), quantifying the concentration of surrounding gas toward core locations. These results suggest that large-scale cloud environments regulate core formation indirectly through the sub-pc-scale intercore medium, where dense cores are assembled.
}


\maketitle

%

\section{Introduction} \label{intro}

The formation of massive stars and stellar clusters is fundamentally a multi-scale process, governed by the hierarchical structure of the interstellar medium (ISM) (e.g., \citealt{Elmegreen2004}, \citealt{McKee2007}, and \citealt{Hennebelle2012}). At galactic scale, molecular gas is first organized into giant molecular filaments (GMFs), extending from several tens of parsecs to a few hundred parsecs and sometimes approaching kilo-pc lengths (e.g., \citealt{Jackson2010}, \citealt{Wang2015, Wang2024}, and \citealt{Hu2026}). These structures may trace spiral-arm features or arise from a combination of galactic dynamics, magnetic turbulence \citep[e.g.,][]{Hu2026}, and feedback-driven compression (e.g., \citealp{Su2026} and \citealp{Bhadari2026}). Continuously linked down to smaller scales \citep{He2026}, clumpy structures as the next hierarchy are called giant molecular clouds (GMCs), which provide molecular reservoirs on scales of $\sim10$--$100$~pc (see reviews in \citealt{Williams2000}, \citealt{Heyer&Dame2015}, and \citealt{Schinnerer2024}). Within GMCs, dense filamentary clouds, ridges and hub--filament systems organize gas on scales of $\sim1$--$10$~pc (e.g., \citealp{Molinari2010a}, \citealp{Andre2014}, \citealt{Kumar2020}, and see review in \citealt{Hacar2023}). These filamentary structures further fragment into pc-scale clumps, which are commonly birthplace of clustered star formation and contain sufficient mass to form groups or clusters of massive stars (see reviews in \citealt{Lada2003}, \citealt{Zinnecker2007} and \citealt{Motte2018}). At still smaller scales, dense cores with characteristic sizes of $\sim0.01$--$0.1$~pc represent the immediate reservoirs for individual or multiple stellar systems, and may be starless, prestellar, or protostellar depending on their gravitational state and embedded source content \citep[see review in][]{Beuther2025a}. Inside protostellar cores, single or multiple stars undergo collapse and accretion process through envelope-disk systems on scales of hundreds to thousands of au (e.g., \citealp{Budaiev2024} and \citealp{Duet1}). This hierarchical sequence, from GMFs and GMCs through dense filaments and clumps, to cores and embedded protostellar disks, provides the physical framework for interpreting how mass is transferred from cloud scales down to the compact structures directly involved in star formation.

Tracing this hierarchy observationally requires both sensitivity to extended emission and sufficient angular resolution to distinguish dense cores. At millimeter wavelengths, single-dish observations reveal the larger gas reservoir, but their angular resolution limits the structures they can resolve. Interferometers such as the Atacama Large Millimeter/submillimeter Array (ALMA) resolve compact emission but filter out spatial scales inadequately sampled by their shortest baselines. This filtering can affect the inferred morphology, masses and column densities (\citealp{Ossenkopf2016} and \citealp{EMERGE-II}). Combining single-dish and interferometric data has been discussed in many works (e.g., \citealp{Stanimirovic2002}, \citealp{Koda2011}, and \citealp{Plunkett2023}). Large-area studies have applied these methods to continuum mosaics (\citealp{GM2013}, \citealp{Battersby2020}, \citealp{Diaz2023}, and \citealp{Ginsburg2026}) and molecular-line mosaics (\citealp{Beuther2022} and \citealp{EMERGE-I}). Mapping large areas at high angular resolution also requires extensive mosaicking and observing time, making cloud-wide studies across these scales challenging.

An evolutionary comparison adds a second requirement to this spatial coverage. The gas distribution and stellar content change as a region progresses from a quiescent, infrared-dark state to active massive star formation and ionized-gas feedback. Measurements confined to one environment therefore provide only a partial view of hierarchical mass assembly. Comparing regions at different stages within the same filament offers a common distance and a shared large-scale setting. Such comparisons can establish how the observed mass distribution differs among environments ordered by their star-formation activity, providing constraints on the evolution of mass concentration.

\begin{figure}
\centering
\includegraphics[width=\linewidth]{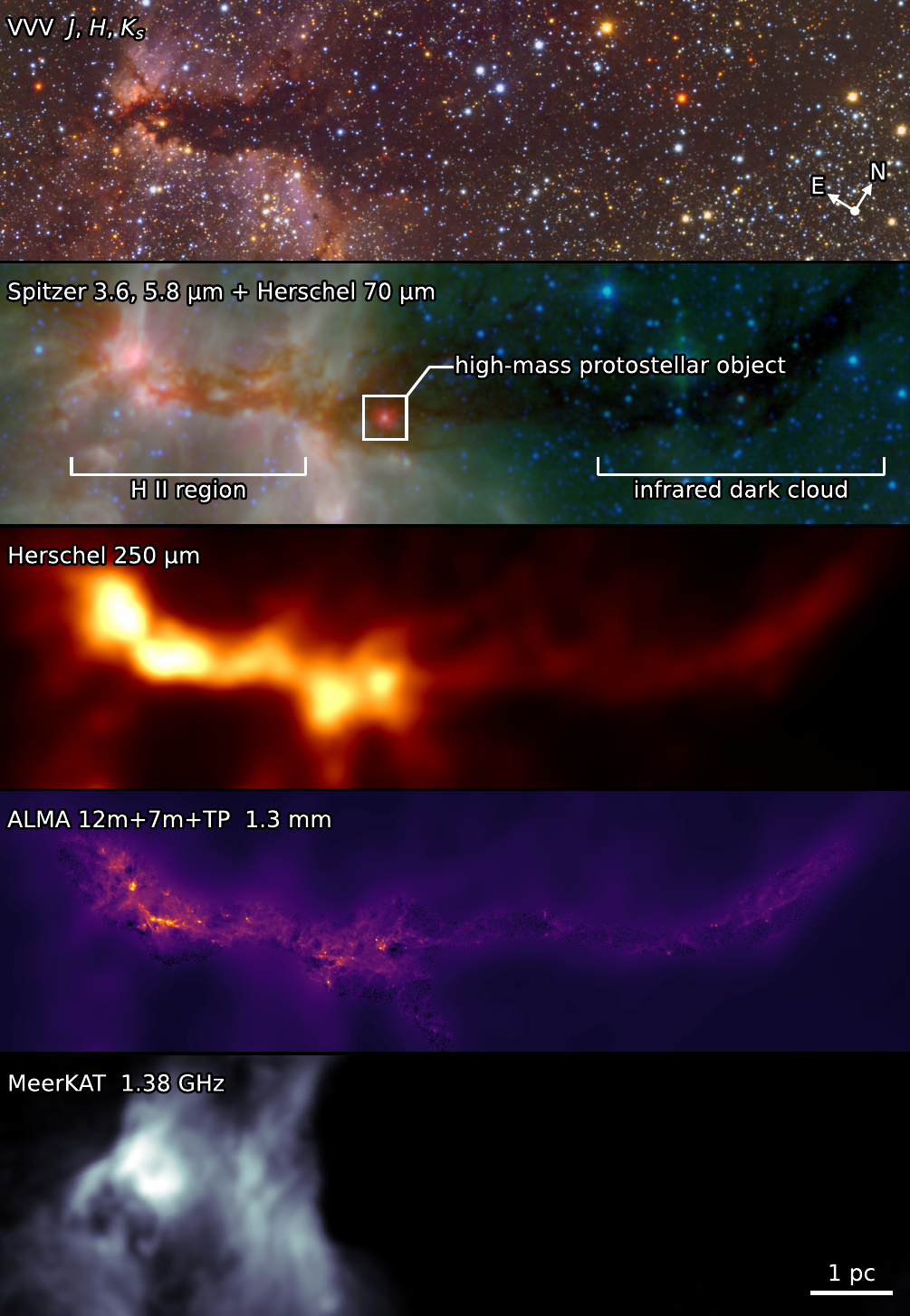}
\caption{\textbf{Multiwavelength view of the G316.8 filament.} From top to bottom: VVV near-infrared composite ($J$, $H$, and $K_s$ in blue, green, and red); Spitzer--Herschel composite (3.6, 5.8, and 70~\um\ in blue, green, and red); Herschel 250~\um; combined ALMA 12~m+7~m+TP 1.3~mm continuum \revise{(inferno colour map, square-root stretch from 0 to 1200~\mjysr; beyond the interferometric mosaic the TP model is shown, cross-faded in where the mosaic primary-beam response falls from 0.5 to 0.2)}; and MeerKAT 1.38~GHz continuum. All strips show the same $14\arcmin\times4\arcmin$ field, rotated along the filament (position angle of 32\degree). The young infrared-dark region is on the right, the intermediate region with a high-mass protostellar object is near the centre, and the evolved H{\sc ii} region is on the left. North and east are indicated in the top strip; the 1~pc scale bar at $D=2.8$~kpc.}
\label{fig:multiwavelength}
\end{figure}

The \textit{Linear Filament and Nested Cluster Evolution Tomography} (LANCET) project\footnote{\url{https://xfengwei.github.io/LANCET/}} was designed to address these both requirements \citep[][hereafter Paper~I]{LANCET-I}. Its target, G316.8, is a massive filament extending over approximately 14~pc. Its heliocentric and Galactocentric distances are $2.8\pm0.5$~kpc and $6.5\pm0.2$~kpc, respectively \citep{Watkins2019}. The filament contains three adjacent regions with contrasting star-formation activity, illustrated by the multiwavelength views in Fig.~\ref{fig:multiwavelength}, allowing the cloud structure and compact sources to be compared within a common observational framework.

Paper~I established the regional segmentation and physical characterization using multiwavelength data. The three subregions are defined as clouds on $\sim4$~pc scales, each hosting a representative clump associated with a different stage of cluster formation. Their boundaries extend beyond these principal clumps to enclose the surrounding material of the parental filamentary clouds. Within each segment, Paper~I characterized the cloud using a molecular hydrogen column-density threshold of $N_{\rm H_2}>3\times10^{22}$~cm$^{-2}$. The resulting cloud masses are $(1.1\pm0.4)\times10^4$, $(1.7\pm0.5)\times10^4$ and $(0.8\pm0.3)\times10^4$~\msun\ for the young, intermediate and evolved regions, respectively. The regions thus encompass the larger parental reservoirs, as well as the representative clumps used to characterize their evolutionary environments.

The evolutionary classification follows the infrared and radio diagnostics summarized in Paper~I. The northern, young region is dark from 3.6 to 70~\um. The central, intermediate region hosts a 70~\um\ source and OH/H$_2$O masers, but lacks detected associated centimeter continuum emission. The southern, evolved region contains a forming OB cluster and H{\sc ii} regions detected through their centimeter continuum emission. The representative clump bolometric luminosity-to-mass ratios, $L_{\rm bol}/M_{\rm clump}$, are 0.7, 26 and 156~\lmsun, respectively, measured on 0.2--0.4~pc scales \revise{(\citealp{Urquhart2022} and \citealp{LANCET-I})}. These clump measurements provide quantitative evolutionary diagnostics for the larger cloud regions in which they are embedded.

In this second LANCET paper, we combine ALMA 12~m and ACA 7~m observations with reconstructed TP information at 1.3~mm. The resulting continuum map connects the $\sim10$-pc filament scale to $\sim0.01$-pc dense-core scales. We ask how the mass transfer efficiencies across cloud ($\sim4$~pc), fragment ($\sim0.15$~pc) and core ($\sim0.01$~pc) scales differ among the three regions. We also examine how core properties relate to the gas between cores and to their broader surroundings. To address these questions, we construct a common core catalogue and separate compact-source emission from the surrounding emission components. After presenting the observations, image combination and catalogue, we examine the mass transfer efficiencies and the relations between cores and their environments.

\section{Observation}\label{sec:obs}

\begin{figure*}[t]
\centering
\includegraphics[width=\textwidth]{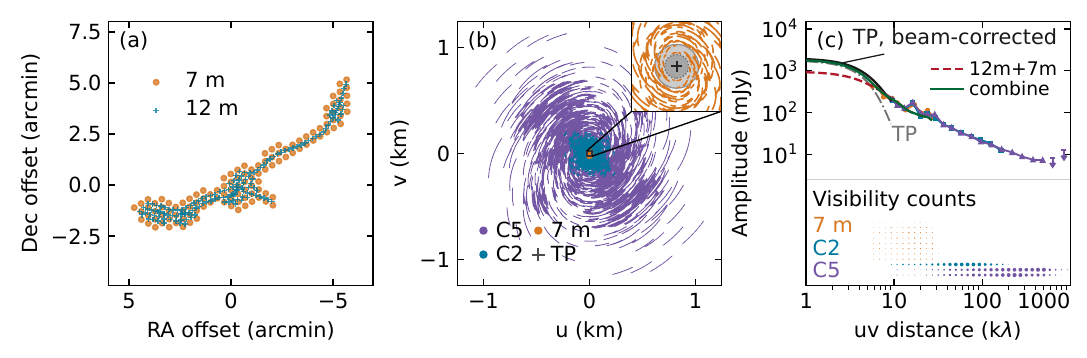}
\caption{\textbf{Continuum pointings and spatial-frequency coverage.} (a) Pointing centres of ACA 7~m (orange circles) and ALMA 12~m C2+C5 (blue crosses), with offsets relative to RA $=221.25^\circ$, Dec $=-59.8^\circ$. (b) Full interferometer $uv$ coverage with axes in km. The central square and black connectors identify the upper-right inset, spanning $-25$ to $+25$~m in both $u$ and $v$ and showing ACA 7~m sampling and TP zero spacing (+). Darker and lighter grey disks mark the 50\% and 10\% Fourier responses of the $18\farcs1$ TP model beam, respectively, at 225~GHz. The legend applies to both the full coverage and the inset. (c) Visibility-amplitude profiles for the 7m (orange), and 12m C2 (blue) and C5 (purple). The 7~m and TP amplitudes are evaluated through the 12~m primary beam. The TP model is shown with its $18\farcs1$ beam (TP) and after removing it (TP, beam-corrected), without flux rescaling. \revise{The dashed red and solid green curves show the amplitudes of the 12~m+7~m image and of the combined image (combine), computed through the same 12~m primary beams and pointing sample as the TP curve with the image restoring beam divided out; they cover 1 to 26~k$\lambda$.} In the lower inset, each row represents an execution block, grouped and coloured by array configuration; vertical positions are categorical, and dot area is proportional to the visibility count in logarithmic $uv$ bins.}
\label{fig:combination_diagnostics}
\end{figure*}

\subsection{ALMA observations}
\label{observe:alma}

The whole G316.8 filament was observed with the mosaic mode in the ALMA project 2016.1.00909.S (PI: Ke Wang). 
As shown in Fig.~\ref{fig:combination_diagnostics}a, 150 pointings of the 12~m array and 118 of the 7~m array were used to cover the filament. The ACA observations from Cycle~4 (Paper~I) and the 12~m observations from Cycle~5 use different mosaic layouts: the ACA pointing centres span a wider area, in addition to the larger field of view of each 7~m dish. Two 12~m array configurations C-5 and C-2 have been used for observations during the ALMA Cycle-5. The C-5 data were obtained using 44 antennas on Jan. 18th and 41 antennas on Mar. 10th in 2018. 
For C-5, the physical baselines range from 15.1 to 1397.8~m, while the projected baselines range from 12.3 to 1241.4~m. The total elapsed time of C-5 configuration is 2.4 hours. The C-2 data were obtained on May 26th in 2018 using 45 antennas. The physical baselines range from 15.0 to 313.7~m, and the projected baselines range from 12.2 to 301.1~m. The total elapsed time of the C-2 configuration is 0.9 hours. For the ACA 7~m array, the physical baselines range from 8.9 to 48.9~m and the projected baselines from 7.0 to 40.9~m. The average on-source integration times per pointing are 36.3~s for C-5, 12.1~s for C-2, and 147~s for the ACA 7~m array. Overlapping pointings increase the effective exposure at a given sky position. Approximating the 225~GHz primary beams as Gaussians with full widths at half maximum of $26\arcsec$ (12~m) and $43\arcsec$ (7~m), and weighting each pointing's integration time by the squared primary-beam response, the median effective exposure at the pointing centres is approximately 85~s for the combined C-5 and C-2 observations and 286~s for the ACA observations. These are approximate sensitivity-equivalent times; they vary across the mosaic and do not account for differences in antenna numbers, flagging between baselines, or imaging weights. Details of ACA 7~m array observations and the spectral setups can be found in Paper~I. With two ALMA 12~m and ACA 7~m arrays, our data are sensitive to a wide range of spatial scales. 

\subsection{Data calibration}
\label{observe:calibrate}

The ACA 7~m data reduction is described in Paper~I, where the self-calibrated visibilities are directly used in this work. The 12~m data were calibrated using the ALMA pipeline within the Common Astronomy Software Applications package \citep[CASA;][]{CASA2022}. The atmosphere, pointing, bandpass, and flux calibrations were performed using J1724$-$4206, while J1524$-$5903 was used as the phase calibrator. Absolute flux calibration uncertainty is $\sim10\%$ from the technical handbook. 

We performed two rounds of phase self-calibration on the combined C-5 and C-2 visibilities, with `inf' and `int' intervals and solution SNR of 3. Line-emission channels were identified using \texttt{findcont} task, visually checked and then flagged. Then the channel-binned continuum visibilities were produced by averaging every 250~MHz bandwidth. The self-calibrated 12~m visibilities were then jointly deconvolved by the \texttt{tclean} task. The clean continuum image is shown as the middle panel in Fig.~\ref{fig:interf_combine}. The representative frequency of continuum is 223.8~GHz or 1.34 mm. The image has a beam size of $0\farcs55 \times 0\farcs36$ with position angle of $50\farcd2$ and reaches a $1\sigma$ root mean square (rms) noise level of $0.1{-}0.2$ \mjybeam, increasing from the northern to the southern fields. There are two reasons for this: 1) the northern fields have more integration time than the southern fields; 2) the rms noise is limited by the dynamic range in the southern fields due to strong sources.

\section{Image combination} \label{sec:combine}

\subsection{ALMA 12~m and ACA 7~m combination} \label{combine:interf}

In Fig.~\ref{fig:combination_diagnostics}b, the $uv$ coverage of 12~m array extends to baselines of $\sim1000$~m, while the ACA 7~m array fills the central $uv$ plane with short-baseline measurements. Their overlap provides feasibility of image combination of two data sets.

Since the two arrays have different primary beam (PB) responses, the effective weighting varies across the field, particularly in large mosaics. In practice, this can lead to ambiguous structures in the dirty images, especially in the regions outside 12~m PB but inside 7~m PB. These features can be difficult to clean robustly, as their flux density and characteristic spatial scales are not well defined by the data alone. If not cleaned or not cleaned correctly, they will behave as residual artifacts inside the PB of the ALMA 12~m. Therefore, standard joint deconvolution by simply co-adding two or multiple interferometric data sets with different PBs can either leave those residual artifacts or require manual and subjective cleaning masks. 

\begin{figure*}
\centering
\includegraphics[width=0.9\linewidth]{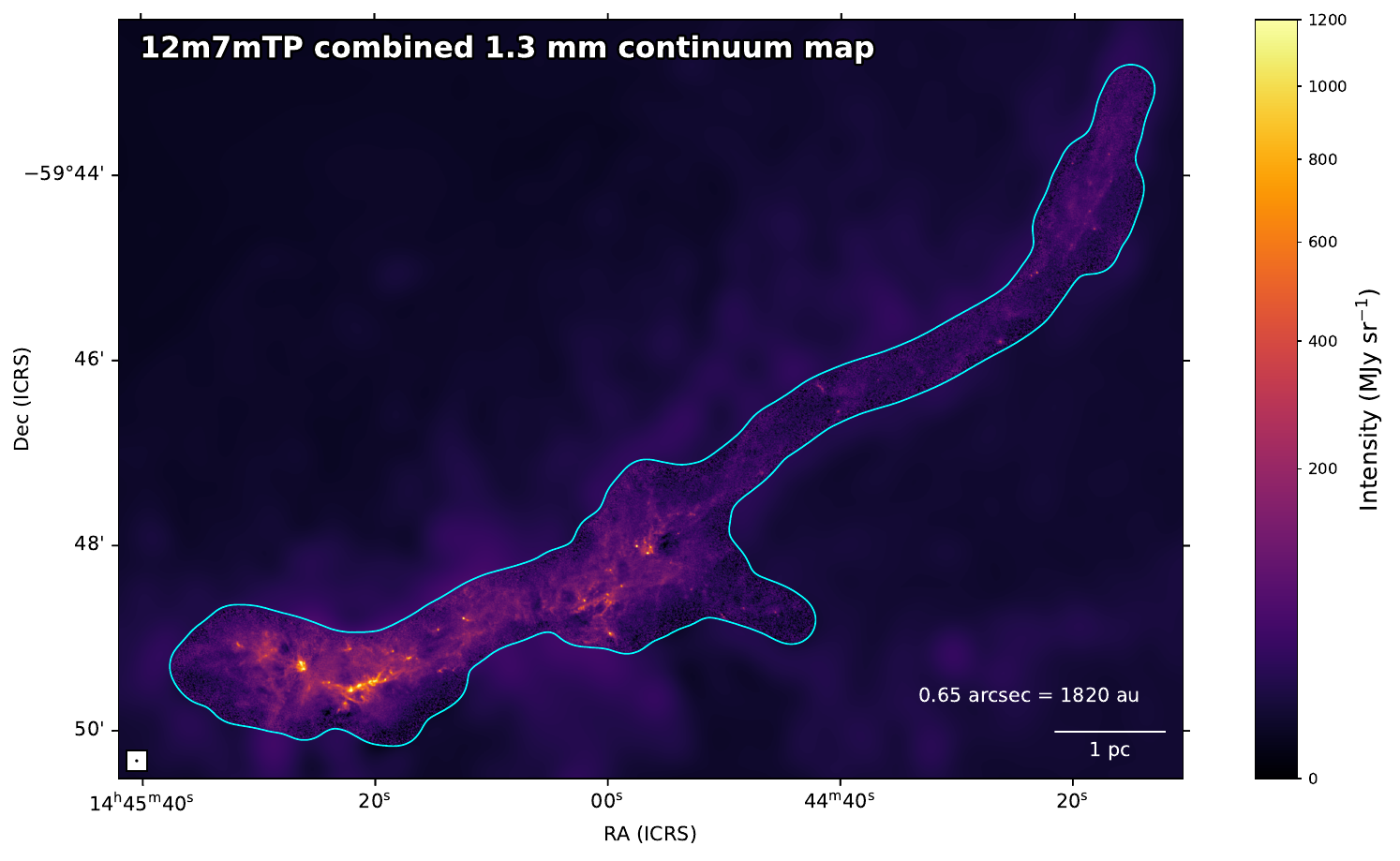}
\caption{{\bf Combined 1.3 mm continuum image.} \revise{The colour scale is a square-root stretch from 0 to 1200~\mjysr. Beyond the coverage of the Paper~I dust model, the TP model is continued with the 18\arcsec\ and 36\arcsec\ SED maps at 36\arcsec\ resolution. The cyan contour marks an effective 12m+7m primary-beam response of 0.2.}}
\label{fig:allscale}
\end{figure*}

To address these challenges, we adopted the MIRIAD-based \citep{MIRIAD} workflow \texttt{almica}\footnote{\url{https://github.com/baobabyoo/almica}} for combination. Here we briefly summarize the key steps of the pipeline. First, ACA 7~m visibilities are deconvolved to produce a clean model. This model is then corrected for the 7~m PB to recover the intrinsic sky brightness distribution. Second, the resulting sky model is attenuated by the 12~m PB. Synthetic 12~m visibilities are then generated for the sky model. Finally, the synthetic visibilities and the original 12-m visibilities are jointly imaged and deconvolved using the combined dirty beam. With this approach, all visibilities share a consistent PB response. The synthetic visibilities are sampled at random $uv$ positions within $\sim9$~m, and their weight relative to the 12~m visibilities is set through an assigned system temperature. Appendix~\ref{app:psf} shows the resulting dirty beam and how the restoring beam is fitted to it. 

Appendix~\ref{app:psf} compares the individual-array and combined images (Fig.~\ref{fig:interf_combine}). The resulting combined image has a synthesized beam of $0\farcs71\times0\farcs61$ at a position angle of $47\farcd4$, corresponding to a physical resolution of $1990~\mathrm{au}\times1710~\mathrm{au}$, or a geometric mean of $1820~\mathrm{au}$, at $D=2.8$~kpc. The rms noise is $\sim80~\mu\mathrm{Jy~beam^{-1}}$, corresponding to a mass sensitivity of $\sim0.01$--$0.03~M_\odot$ per beam for dust temperatures of 10--20~K. 

\subsection{Interferometric and total-power combination} \label{combine:interfTP}

\begin{figure*}
\centering
\includegraphics[width=0.9\linewidth]{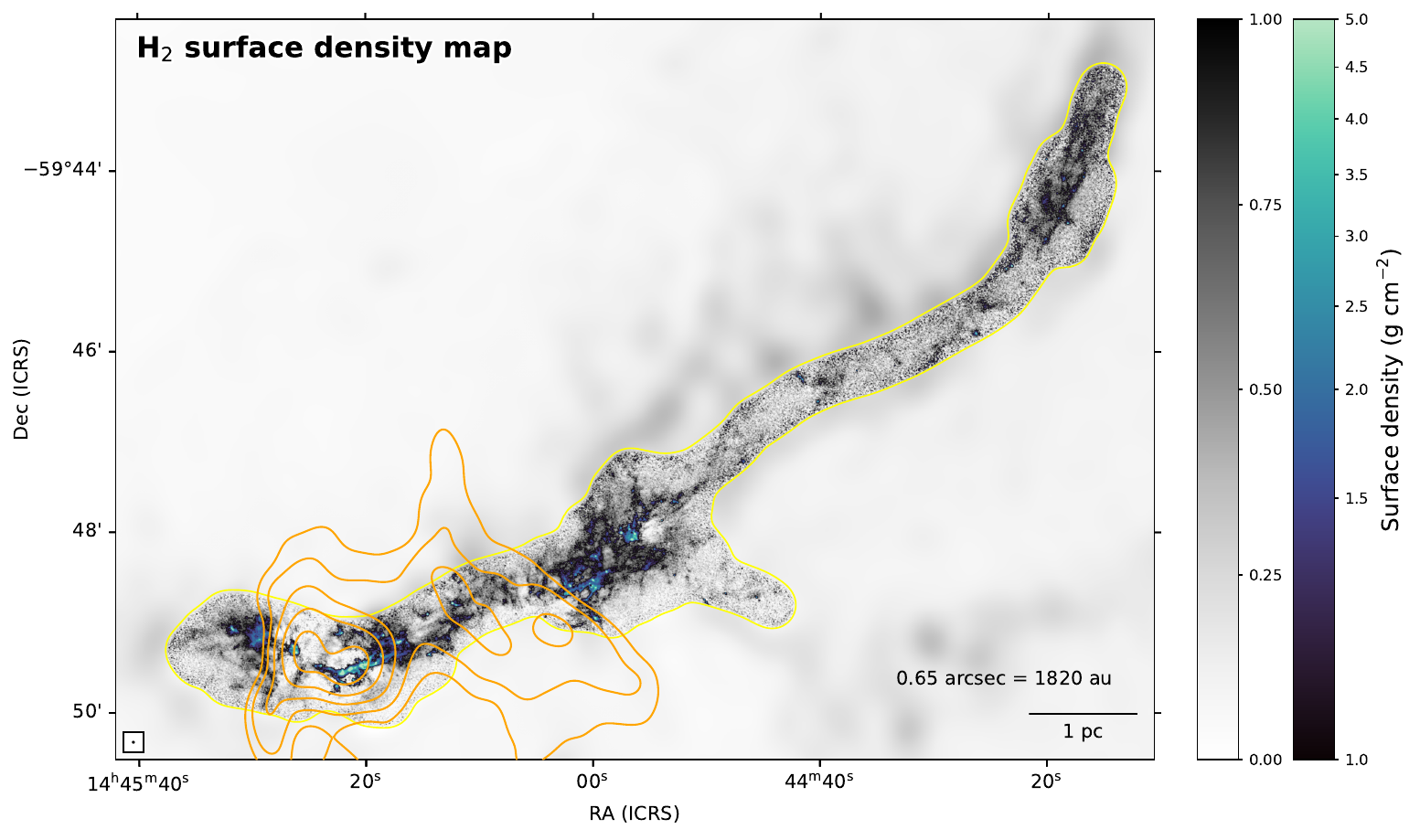}
\caption{{\bf Molecular hydrogen surface density map.} Dual color bars are used. The first one (in gray scale) linearly starts from 0 to \revise{1}~\gpsc, and the second one (in color scheme) starts from \revise{1} and saturates at \revise{5}~\gpsc\ in a square-root stretch. \revise{Pixels with negative intensity are shown at their absolute value.} Free-free emission is shown in orange contours with levels of 0.05, 0.1, 0.2, 0.4, and 0.8~\jybeam. The yellow contour marks the effective 12m+7m primary-beam response of 0.2. One can convert this map to H$_2$ number column density by a factor of $2.14\times10^{23}~\mathrm{g}^{-1}$.}
\label{fig:sigma}
\end{figure*}

Our TP 1.3 mm continuum data, constructed from dust and free-free spectral energy distribution (SED) fitting (Appendix~\ref{app:TP}), have $\sim18\arcsec$\ angular resolution, or an effective diameter of $\sim22$~m. The inset in Fig.~\ref{fig:combination_diagnostics}b shows the ACA short-baseline sampling together with the Fourier response of the TP model beam. We applied the J-comb algorithm \citep{Jiao2022} to linearly combine the PB-corrected \texttt{12m7m} and TP images. \revise{The TP model enters the combination at its own flux scale, without rescaling. We checked this scale against the Planck 217~GHz map, scaled to 223.8~GHz after removing the CO~(2--1) contribution: at the $5\arcmin$ Planck resolution, where the TP model covers at least 90\% of the beam, the ratio TP/Planck is 1.13 (16th--84th percentiles 0.98--1.32) after removing a uniform line-of-sight offset of $\sim4$~\mjysr.} Compared to feathering, J-comb uses a different spatial-frequency weighting and tapering scheme designed to preserve a nearly Gaussian beam response in the combined image. The resulting 1.3~mm continuum image (Fig.~\ref{fig:allscale}) connects compact structures to the extended filament emission restored by the TP model. Together with the dust surface-density map (Fig.~\ref{fig:sigma}), it shows the cores, filamentary structures, hubs and cavities in their cloud-scale context. Compared with the interferometric-only image in Fig.~\ref{fig:interf_combine}, the combined image recovers more diffuse emission and mitigates negative bowls.

We produced another version of the combined image using only the TP dust thermal emission component (Fig.~\ref{fig:synthetic}, left panel). Assuming single-component dust emission along the line of sight, the surface density map is derived by
\begin{equation} \label{eq:surfden}
\Sigma = \frac{-\ln\left[1 - I^{\rm dust}_{\nu} / B_\nu(T_{\rm dust}) \right]}{\kappa_{\nu}/R_{\rm gd}},
\end{equation}
where $I^{\rm dust}_{\nu}$ is the combined 1.3 mm dust continuum intensity, $B_{\nu}(T_{\rm dust})$ is the Planck function at dust temperature $T_{\rm dust}$. We assume that the small-scale dust temperature is equal to the environmental temperature ($T_{\rm env}$) at an angular resolution of 18\arcsec, as derived from the SED analysis (given in Paper~I). We adopt the dust opacity law $\kappa_{\nu} = 10\times(\nu[\mathrm{GHz}]/10^3)^\beta$~\scpg{} with $\beta=2$ \citep{Hildebrand1983}, which gives $\kappa_{\nu}=0.5$~\scpg. We obtain the gas-to-dust ratio $R_{\rm gd}=101\left[^{+12}_{-30}\right] \pm 13$ from the statistical correlation \citep{Giannetti2017}. The uncertainty budget of surface density comes from gas-to-dust ratio ($24\%$), dust temperature assumption ($20\%$), and absolute flux ($10\%$), resulting in a relative uncertainty of 33\%. The surface density map is shown in Fig.~\ref{fig:sigma}.

Clear morphological and structural variations are seen from north to south. The young region exhibits a globally low density contrast (\revise{mostly 0.3--0.8~\gpsc}), with isolated compact sources locally \revise{reaching 3--13~\gpsc}. \revise{Elongated and narrow dusty filaments (recently discovered in high-resolution dust emission, \citealp{Zhang2026}) are largely aligned with the main filament of G316.8, and compact sources are located primarily at the intersections of these fibrils}\revise{, as expected if cores form where colliding subfilaments merge \citep{Li2026}}. In the intermediate region, high-density gas becomes more widespread\revise{: the area above 1~\gpsc\ rises from 6\% in the young region to 10\% in the intermediate and 13\% in the evolved region, as fractions of the sky area of each subregion}. Extended, filamentary structures remain prominent, but they show a more converging pattern, possibly reflecting a hub--filament configuration inherited from larger scales. In contrast, the most evolved region is dominated by a prominent bipolar cavity oriented perpendicular to the filament. This structure is likely carved by the bipolar H{\sc ii} region identified in Fig.~1 of Paper~I. As a comparison, this cavity is filled by free-free emission from ionized gas in Fig.~\ref{fig:sigma}. 

\section{Continuum compact source catalogue} \label{sec:cat}

\subsection{Source and background decomposition}
\label{cat:extract}

We adopt the \texttt{getsf} algorithm \citep{getsf} that can identify sources and filaments simultaneously to decompose and characterize compact, roundish sources (dense cores) from complex protocluster environment. It enables background subtraction and multi-scale/image processing, and has been applied to many studies of cores already (e.g., ALMA-IMF, \citealp{Motte2022}; ASSEMBLE, \citealp{Xu2024Assemble1}; ALMA-QUARKS, \citealp{Jiao2026}). In the test field (Appendix~\ref{app:getsf}), including the 7~m data increases the number of detected sources and changes the measured fluxes, while adding TP has a smaller effect on most compact-source measurements. The reconstructed TP contribution restores extended emission in the \texttt{getsf} background, providing the cloud-scale context and the foreground/background estimate used in the radiative-transfer analysis. This behavior is consistent with the scale dependence of interferometric filtering, which preferentially retains compact structure while attenuating extended emission \citep{Miao&Li2026}. In this work, we adopt the \texttt{getsf} extraction in both {\tt 12m7mTP} and {\tt 12m7m} images and combine them for constructing the compact source catalogue. 

\begin{figure}
\centering
\includegraphics[width=0.8\linewidth]{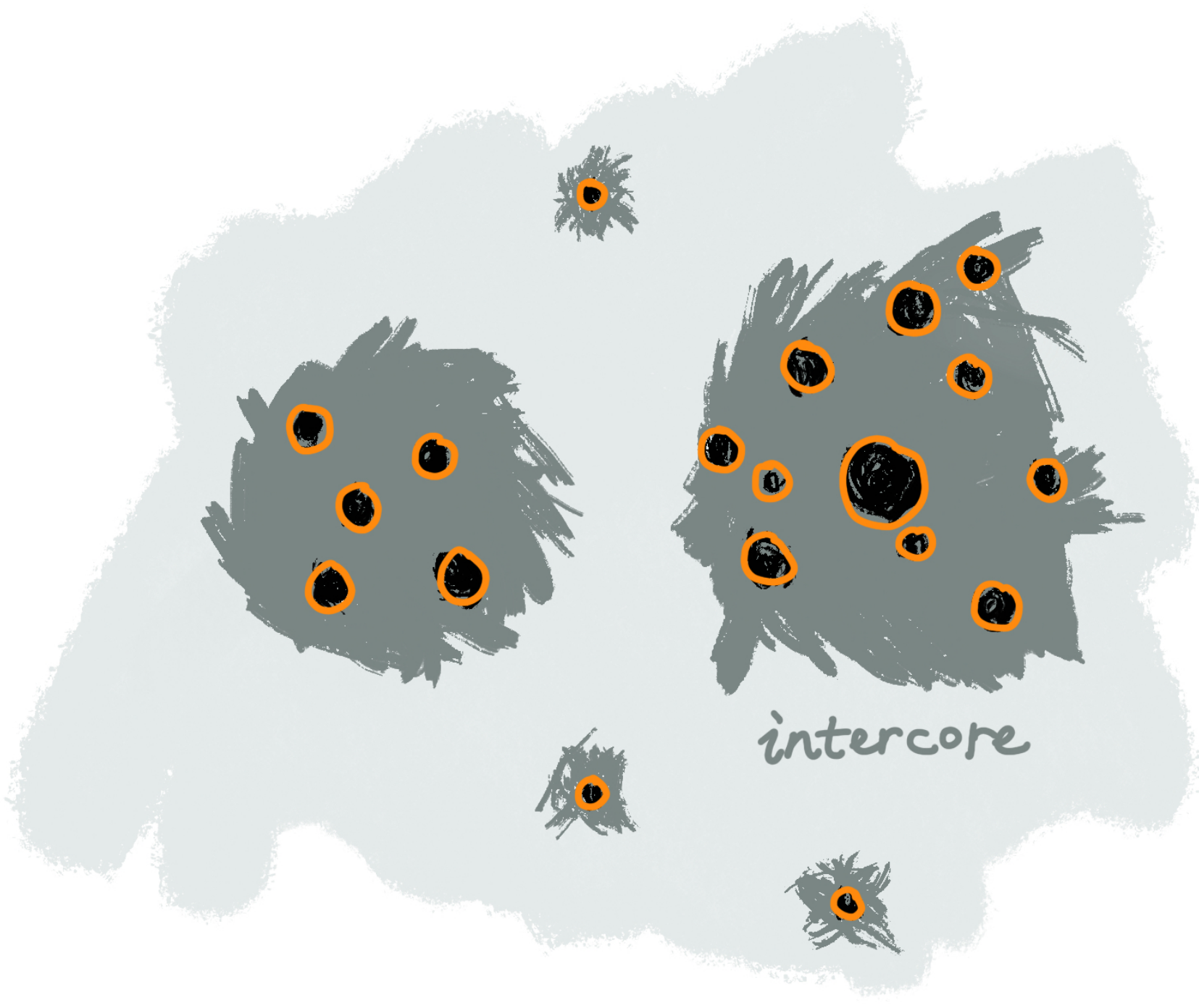}
\caption{{\bf Schematic demonstration of source and background decomposition.} Black dots with orange outlines indicate dense cores while dark gray shades are so-called intercore medium which are the material surrounding dense cores within pc scale. The light gray shade is cloud-scale emission that is even more extended than pc.}
\label{fig:intercore}
\end{figure}

For source extraction, we multiplied the PB-corrected \texttt{12m7m} and \texttt{12m7mTP} images by the same effective 12m+7m mosaic gain, $\mathcal{G}(x,y)$, to obtain a more nearly uniform rms across the field. For \texttt{12m7mTP}, this attenuation is applied after J-comb to the combined image within, including the TP contribution. We restricted extraction to $\mathcal{G}>0.2$, whose boundary is shown in Figs.~\ref{fig:allscale} and \ref{fig:sigma}. The source and background components were decomposed for both the \texttt{12m7mTP} and \texttt{12m7m} images, with an illustrative example shown in Fig.~\ref{fig:sbackground}. This decomposition provides a practical way to separate the multiscale components traced by the continuum emission: the compact-source component corresponds to dense cores, the source-removed \texttt{12m7m} background traces the dust emission distributed between dense cores, and the additional extended emission recovered only after including TP data represents the cloud-scale material. As illustrated in Fig.~\ref{fig:intercore}, we refer to the source-removed \texttt{12m7m} background as the intercore medium, denoted as $I_{\nu}^{\rm ic}$. This terminology is motivated by the classical picture of clumpy molecular clouds, in which dense clumps or cores are embedded within a more diffuse inter medium component \citep{Stutzki&Guesten1990}. In this work, however, the term is used operationally: $I_{\nu}^{\rm ic}$ represents the dust emission remaining between compact dense cores after source removal, while structures larger than the maximum recoverable scale of the \texttt{12m7m} image, $\sim28\arcsec$\ ($\sim0.4$~pc), are filtered out. In contrast, the background component of the \texttt{12m7mTP} image, $I_{\nu}^{\rm bg}$, preserves the large-scale flux after compact sources are removed, thereby tracing the combined contribution from the intercore medium and the TP-recovered cloud-scale material.

\subsection{Source measurement and evaluation}
\label{cat:measure}

\begin{figure*}
\centering
\includegraphics[width=\linewidth]{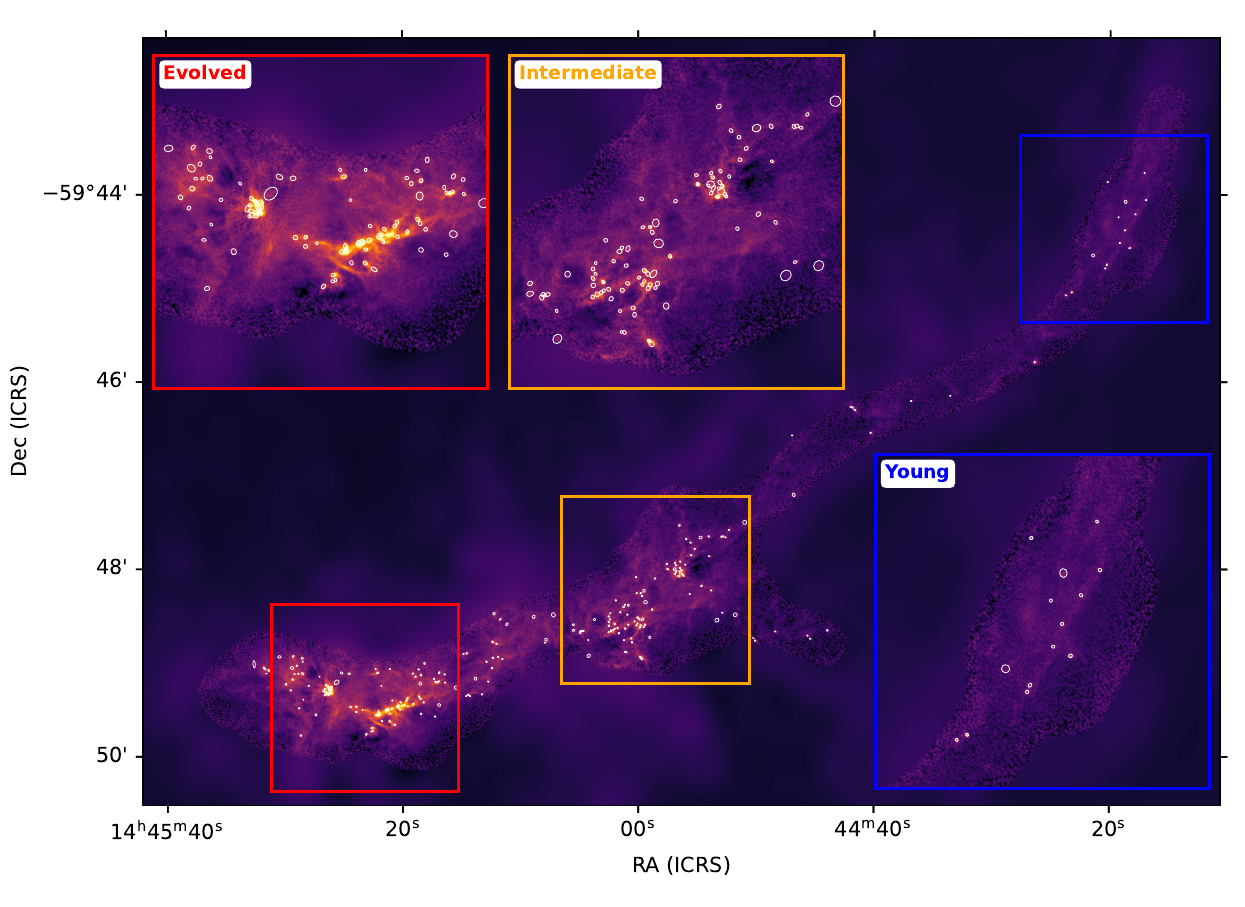}
\caption{{\bf Compact source extraction and spatial distribution.} Color map displays the \texttt{12m7mTP} 1.3 mm \revise{continuum emission (dust and free--free), on a square-root scale from 0 to 1000~\mjysr}. The main panel provides an overview of the filament, with extracted compact sources marked with white ellipses. Three representative subregions (each with size of $2'\times2'$) at different evolutionary stages are zoomed within the inset panels. In the zoomed-in panels, the ellipse sizes are scaled by a factor of 1.5 relative to their measured values for clarity.}
\label{fig:getsf_map}
\end{figure*}

\begin{figure*}
\centering
\includegraphics[height=0.35\linewidth]{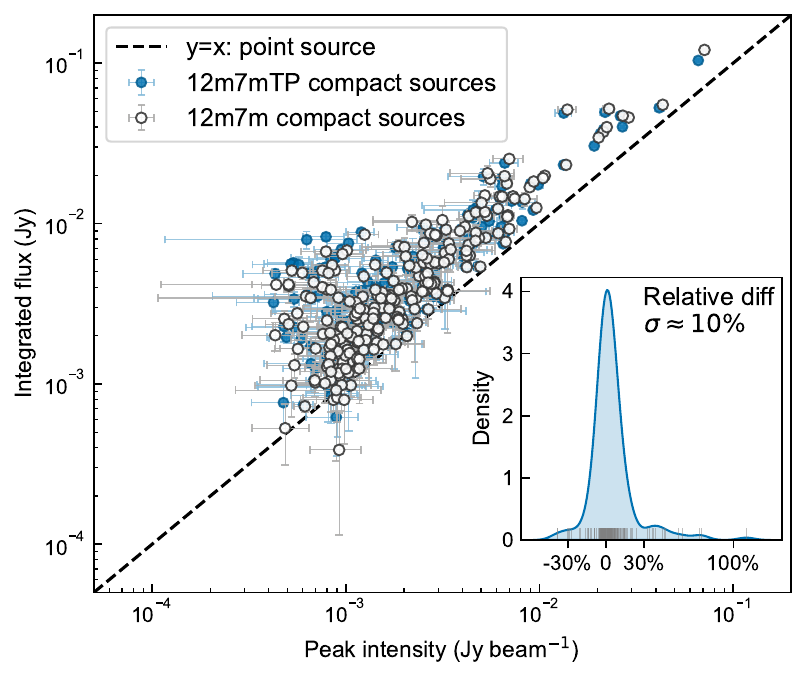}
\includegraphics[height=0.35\linewidth]{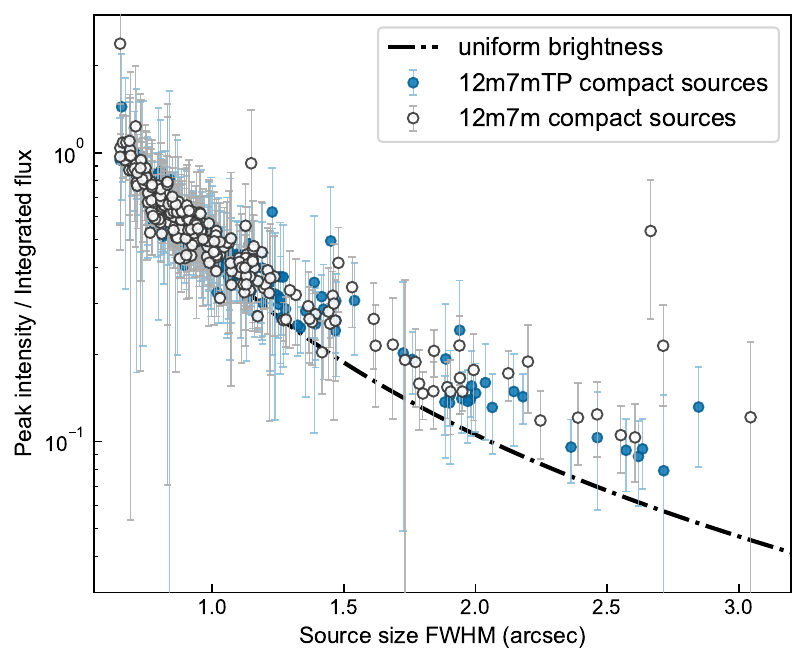}
\caption{{\bf Evaluation on source measurements.} {\it Left}: Measured total flux versus peak intensity. The secondary panel indicates the relative difference in flux density between sources measured in two images, which is about 10\%. The dashed line marks the point-source limit, where the peak intensity equals the integrated flux. {\it Right}: Ratio of total flux over peak intensity as a function of source FWHM. The dash-dotted curve represents the case of a source with uniform surface brightness, for which the ratio decreases with the square of the source size. Orange color is for {\tt 12m7mTP} while white is for {\tt 12m7m} measurements. }
\label{fig:getsf_flux}
\end{figure*}

In total, 231 sources are identified. Their spatial distribution is shown in Fig.~\ref{fig:getsf_map}. All the detections are listed in Table~\ref{tab:getsf}, which are sorted by descending signal-to-noise ratio. The peak intensity and integrated flux densities, together with their associated uncertainties, are obtained from the elliptical Gaussian fitting. These flux-related quantities, denoted as $\mathcal{F}$, are divided by the same gain $\mathcal{G}(x,y)$ at each source position to remove the attenuation introduced for extraction and recover PB-corrected photometry,
\begin{equation}
\mathcal{F}_{\rm pbcor} = \mathcal{F} / \mathcal{G}(x,y).
\end{equation}
We assess the consistency of the flux measurements between the {\tt 12m7m} and {\tt 12m7mTP} images. As shown by the orange and white data points in Fig.~\ref{fig:getsf_flux}, the distribution of relative flux differences is approximately Gaussian, with a data scatter of $\sigma \sim 10\%$. A small number of outliers exhibit {\tt 12m7mTP} fluxes more than 30\% higher than those measured from {\tt 12m7m}. We have checked that these cores have relatively flat surface density radial profiles, where the TP data can make a significant contribution to the recovered flux. In the following, we therefore adopt a systematic flux uncertainty of 10\% to account for the dependence on source extraction strategy. Combining with 10\% absolute flux uncertainty at Band~6 for ALMA, the uncertainty of source flux measurement is about 14\% since the two uncertainties are independent.

To assess the reliability of the flux measurements and the degree to which sources are spatially resolved, we examine the relationship between integrated flux, peak intensity, and observed source size (see Fig.~\ref{fig:getsf_flux}). The left panel shows that all except for seven of the sources lie above the one-to-one relation expected for point sources, indicating that they are not point sources. For the remaining seven sources, their signal-to-noise ratios are all smaller than 10. So their flux measurements can be easily affected by noise or interferometric artifacts, like negative values close to source edge. The right panel of Fig.~\ref{fig:getsf_flux} further demonstrates that the ratio of peak intensity to integrated flux decreases with increasing source size, as expected for spatially resolved emission. The uniform-brightness curve is drawn to indicate those sources of which the ratio decreases with the square of the source size. The curve starts from the beam size in x axis and ratio of 1 in y axis. We find that $>80\%$ sources lie above the uniform-brightness curve. Such a trend implies that their emission is more centrally peaked rather than uniformly distributed. 

\begin{table*}
\centering
\caption{{\bf Measurements of compact sources in the G316.8 filament} \label{tab:getsf}}
\begin{tabular}{ccccccccccccc}
\hline
\hline \noalign{\smallskip}
ID & R.A.(2000) & Dec.(2000) & $\mathcal{R}$ & S/N & $\theta^{\rm obs}_{\rm maj}\times\theta^{\rm obs}_{\rm min}$ & $\mathrm{PA}_{\rm obs}$ & $\theta_{\rm src}$ & Res. & $I^{\rm peak}_\nu$ & $S^{\rm tot}_\nu$ & $I^{\rm bg}_\nu$ & $I^{\rm ic}_\nu$ \\
 & ($h$:$m$:$s$) & ($^\circ$:$'$:$''$) &  &  & ($\arcsec\times\arcsec$) & (deg) & (\arcsec) & & (\mjybeam) & (mJy) & \multicolumn{2}{c}{(\mjysr)} \\
\hline \noalign{\smallskip}
1 & 14:45:26.38 & -59:49:15.47 & E & 545 & 0.92$\times$0.71 & 32 & 0.48 & Y & 71.2(1.4) & 121.1(1.5) & \revise{696} & 696 \\
2 & 14:45:26.13 & -59:49:19.62 & E & 333 & 0.78$\times$0.65 & 64 & $\leq0.33$ & N & 43.4(1.9) & 55.1(1.8) & \revise{742} & 730 \\
3 & 14:44:56.59 & -59:48:04.71 & I & 241 & 0.86$\times$0.75 & 74 & 0.47 & Y & 27.0(1.0) & 47.1(1.3) & \revise{341} & 296 \\
4 & 14:44:59.85 & -59:48:56.48 & I & 235 & 0.80$\times$0.74 & 60 & 0.41 & Y & 28.9(0.6) & 45.9(0.8) & \revise{240} & 204 \\
5 & 14:45:12.40 & -59:48:46.73 & I & 206 & 0.87$\times$0.84 & 53 & 0.56 & Y & 22.3(0.6) & 40.0(0.7) & \revise{199} & 132 \\
6 & 14:44:57.53 & -59:48:00.17 & I & 193 & 0.87$\times$0.80 & 7 & 0.52 & Y & 21.4(0.4) & 37.3(0.5) & \revise{165} & 114 \\
7 & 14:45:22.07 & -59:49:33.53 & E & 186 & 1.03$\times$0.83 & 2 & 0.65 & Y & 22.9(1.5) & 52.0(1.8) & \revise{739} & 691 \\
8 & 14:45:17.10 & -59:49:12.49 & E & 162 & 0.81$\times$0.78 & 161 & 0.46 & Y & 20.2(0.5) & 34.4(0.7) & \revise{291} & 202 \\
9 & 14:45:02.02 & -59:48:35.34 & I & 131 & 0.89$\times$0.69 & 40 & 0.43 & Y & 13.7(0.6) & 23.2(0.7) & \revise{236} & 181 \\
10 & 14:44:58.84 & -59:48:26.10 & I & 103 & 0.96$\times$0.78 & 15 & 0.56 & Y & 9.4(0.4) & 18.3(0.5) & \revise{166} & 90 \\
\hline \\
\end{tabular}
\tablefoot{
ID: Source index ordered by decreasing signal-to-noise ratio (S/N). R.A. and Dec.: Right ascension and declination. $\mathcal{R}$: Subregion labels, Y for young, I for intermediate, and E for evolved. $\theta^{\rm obs}_{\rm maj}\times\theta^{\rm obs}_{\rm min}$: Observed major and minor FWHM angular sizes. $\mathrm{PA}_{\rm obs}$: Position angle of the observed ellipses, measured east of north. $\theta_{\rm src}$: Intrinsic angular sizes after deconvolution. Res.: Marker of whether source is resolved or not. N denotes a size upper limit; the treatment of density limits is described in Sect.~\ref{cat:property}. $I^{\rm peak}_\nu$: Peak intensity, as the flux density over beam solid angle at the brightest pixel. $S^{\rm tot}_\nu$: Integrated flux density over source solid angle. $I^{\rm bg}_\nu$: \texttt{getsf} separated background brightness from the {\tt 12m7mTP} image. $I^{\rm ic}_\nu$: intercore background brightness from the \texttt{getsf} source-removed {\tt 12m7m} image, sampled at the core peak. Only the first ten rows are shown and the complete version is published online. 
}
\end{table*}

G316.8 has three distinct evolutionary subregions as defined in Sect.~\ref{intro}. Based on spatial association with evolutionary subregions, sources are classified: there are 16 in young (Y), 120 in intermediate (I), and 95 in evolved (E) parts, respectively. Note that this is not the evolutionary stages of sources themselves, but only presents the stages of their parental clouds, or the representative evolution of environments.

\subsection{Physical properties of dense cores} \label{cat:property}

We estimate intrinsic FWHM by circularized beam subtraction,
\begin{equation}
\theta_{\rm src}=\left(\theta^{\rm obs}_{\rm maj}\theta^{\rm obs}_{\rm min}-\theta_{\rm b,maj}\theta_{\rm b,min}\right)^{1/2}.
\label{eq:size_deconvolution}
\end{equation}
A source is flagged as resolved only when $\theta_{\rm obs}^{2}\geq1.25\theta_{\rm b}^{2}$, where $\theta_{\rm obs}^{2}=\theta^{\rm obs}_{\rm maj}\theta^{\rm obs}_{\rm min}$ and $\theta_{\rm b}^{2}=\theta_{\rm b,maj}\theta_{\rm b,min}$. The remaining 22 of 231 sources (9.5\%), flagged N in Table~\ref{tab:getsf}, are assigned the upper size limit $\theta_{\rm src}\leq\theta_{\rm b}/2=0.326\arcsec$, corresponding to approximately 900~au. Their size-based volume densities are lower limits because, at fixed mass, the inferred density scales as the inverse cube of the source size. We use the 209 resolved cores for size and volume-density statistics and correlations involving $n_{\rm core}$. Beam-averaged surface densities, integrated masses and environmental measurements retain all valid sources.
Source physical size was derived from its angular size by
\begin{equation} \label{eq:lsrc}
\ell_{\rm src} = D\cdot \theta_{\rm src}
\end{equation}
where $D=2.8~\mathrm{kpc}$ is the distance to the source and $\theta_{\rm src}$ is the deconvolved source size. For the 209 resolved sources, $\ell_{\rm src}$ ranges from approximately 920 to 8300~au, with a median of 2000~au. We call them dense cores hereafter.

\begin{table*}
\centering
\caption{{\bf Physical properties of dense cores in the G316.8 filament} \label{tab:physics}}
\begin{tabular}{cccccccccc}
\hline
\hline \noalign{\smallskip}
ID & $T_{\rm env}$ & $\ell_{\rm src}$ & $\Sigma^{\rm ic}_{\rm core}$ & $\Sigma^{\rm bg}_{\rm core}$ & $\tau^{\rm c}_{\nu}$ & $\Sigma_{\rm core}$ & $M_{\rm core}$ & $\langle \rho_{\rm core} \rangle$ & $\langle n_{\rm core} \rangle$ \\
 & (K) & (au) & (\gpsc) & (\gpsc) &  & (\gpsc) & (\msun) & ($10^{-16}~\mathrm{g~cm}^{-3}$) & ($10^{8}~\mathrm{cm}^{-3}$) \\
\hline \noalign{\smallskip}
1 & 33.0 & 1300 & 3.6$\pm$1.4 & \revise{3.6$\pm$1.4} & \revise{0.163$\pm$0.023} & 35$\pm$14 & 26$\pm$13 & 15$\pm$7 & 3.3$\pm$1.5 \\
2 & 30.3 & $\leq900$ & 4.1$\pm$1.6 & \revise{4.2$\pm$1.7} & 0.107$\pm$0.015 & 23$\pm$9 & 12$\pm$7 & $\geq24$ & $\geq5.0$ \\
3 & 19.3 & 1300 & 2.9$\pm$1.2 & \revise{3.4$\pm$1.4} & 0.117$\pm$0.018 & 25$\pm$10 & 19$\pm$10 & 12$\pm$6 & 2.6$\pm$1.2 \\
4 & 22.5 & 1100 & 1.6$\pm$0.7 & \revise{1.9$\pm$0.8} & 0.102$\pm$0.015 & 22$\pm$9 & 15$\pm$8 & 15$\pm$7 & 3.1$\pm$1.4 \\
5 & 20.7 & 1600 & 1.2$\pm$0.5 & \revise{1.8$\pm$0.7} & \revise{0.086$\pm$0.013} & 19$\pm$8 & 14$\pm$8 & 5.5$\pm$2.5 & 1.2$\pm$0.5 \\
6 & 18.5 & 1500 & 1.2$\pm$0.5 & \revise{1.7$\pm$0.7} & \revise{0.096$\pm$0.014} & \revise{21$\pm$8} & 16$\pm$8 & 7$\pm$3 & 1.5$\pm$0.7 \\
7 & 24.4 & 1800 & 5.1$\pm$2.0 & \revise{5.5$\pm$2.2} & 0.073$\pm$0.011 & 16$\pm$6 & 15$\pm$8 & 3.6$\pm$1.7 & 0.8$\pm$0.4 \\
8 & 20.0 & 1300 & 1.9$\pm$0.8 & \revise{2.7$\pm$1.1} & \revise{0.082$\pm$0.012} & 18$\pm$7 & 13$\pm$7 & 9$\pm$4 & 1.9$\pm$0.9 \\
9 & 19.5 & 1200 & 1.8$\pm$0.7 & \revise{2.3$\pm$0.9} & 0.057$\pm$0.009 & 12$\pm$5 & 9$\pm$5 & 7$\pm$3 & 1.5$\pm$0.7 \\
10 & 18.6 & 1600 & 0.9$\pm$0.4 & \revise{1.7$\pm$0.7} & 0.041$\pm$0.006 & 9$\pm$4 & 7$\pm$4 & 2.7$\pm$1.3 & 0.58$\pm$0.27 \\
\hline \\
\end{tabular}
\tablefoot{
ID: Source index ordered by decreasing signal-to-noise ratio (Sig). $T_{\rm env}$: Environmental dust temperature sampled at the core peak. $\ell_{\rm src}$: Intrinsic source physical size in unit of au. $\Sigma^{\rm ic}_{\rm core}$: Intercore surface density at the core peak, derived from $I^{\rm ic}_\nu$. $\Sigma^{\rm bg}_{\rm core}$: Surface density summing up foreground and background material along the line of sight. $\tau^{\rm c}_{\nu}$: Dust optical depth of dense core at 1.3 mm. $\Sigma_{\rm core}$: Surface density of dense core. $M_{\rm core}$: Mass of dense core. $\langle \rho_{\rm core} \rangle$: Mean volume density of hydrogen molecules of dense core. $\langle n_{\rm core} \rangle$: Mean number volume density of hydrogen molecules of dense core. For unresolved sources flagged N, sizes are upper limits and volume and number densities at fixed mass are lower limits. These limits are excluded from the size and volume-density statistics and correlations. The listed $\tau^{\rm c}_{\nu}$ and $\Sigma_{\rm core}$ are beam-averaged peak estimates, independent of the source-size limit. Only the first ten rows are shown and the complete version is published online.
}
\end{table*}

We calculate the core optical depth $\tau^{\rm c}_\nu$ using the radiative-transfer equation through the foreground, dense core, and background (Eq.~\ref{eq:tau_c}). The TP-containing image provides the large-scale foreground and background emission needed in this calculation. For the adopted core temperatures and peak intensities, including this emission increases the inferred core optical depth by a median of \revise{0.7\%} and at most \revise{2.8\%} compared with setting the foreground and background emission to zero. Appendix~\ref{app:rt} illustrates how the effect increases for brighter backgrounds. Core mass, corrected by optical-depth effect, is written as
\begin{equation}
\begin{split}
M_{\rm core}
& = M_{\rm thin} \times f_\tau \\
& = \frac{S^{\rm tot}_\nu\, d^2 R_{\rm gd}}{\kappa_\nu B_\nu (T_{\rm c})} \times \frac{\tau^{\rm c}_\nu}{1-e^{-\tau^{\rm c}_\nu}}
\end{split},
\end{equation}
where $M_{\rm thin}$ is core mass for optically thin dust thermal emission, and $f_{\tau}$ is optical-depth correction factor, with a maximum value of $1.084$ in our case. The dust opacity $\kappa_\nu$, gas-to-dust ratio $R_{\rm gd}$, and Planck function $B_\nu(T_{\rm c})$ are the same as in Eq.~\ref{eq:surfden}. The core temperature $T_{\rm c}$ is assumed to equal the environmental dust temperature $T_{\rm env}$ sampled at the core peak from the SED-fitted temperature map of Paper~I. The obtained core masses range from 0.12 to 26~\msun, with a median value of 1.4~\msun. 

The surface density of dense core is obtained by
\begin{equation} \label{eq:tau2surf}
\Sigma_{\rm core} = (\tau^{\rm c}_{\nu} / \kappa_{\nu})\cdot R_{\rm gd}
\end{equation}
where $\kappa_{\nu}$ and $R_{\rm gd}$ have the same value as explained in Eq.~\ref{eq:surfden}. The beam-averaged surface densities of all 231 cores range from 0.25 to 35~\gpsc, with a median of 1.3~\gpsc.

We calculate two surface densities of different background products as introduced in Sect.~\ref{cat:extract}. $\Sigma^{\rm bg}_{\rm core}$, defined as summation of surface density of foreground and background medium towards dense cores, is obtained by substituting opacity $\tau_{\nu}^{\rm fb}$ via radiation transfer solution from Eq.~\ref{eq:tau_fb} into Eq.~\ref{eq:tau2surf}. Its values range from \revise{0.3 to 6.0}~\gpsc, with a median value of \revise{1.4}~\gpsc. $\Sigma^{\rm ic}_{\rm core}$ is the local intercore surface density, calculated by substituting the source-removed brightness $I^{\rm ic}_\nu$ sampled at the core peak into Eq.~\ref{eq:surfden}. It describes material around the core and is distinct from the compact-core surface density $\Sigma_{\rm core}$. Its values range from 0.1 to 5.7~\gpsc, with a median value of 0.8~\gpsc. 

Assuming spherical symmetry, the mean volume density of a dense core can be written as
\begin{equation} \label{eq:volden}
\langle n_{\rm core}\rangle = \frac{\langle \rho_{\rm core} \rangle}{\mu m_{\rm H}} = \frac{3M_{\rm core}}{4\pi \mu m_{\rm H} \ell_{\rm src}^3},
\end{equation}
where $\mu=2.8$ is the mean molecular weight per hydrogen molecule and $m_{\rm H}$ is the mass of a hydrogen atom. Following \citet{Rosolowsky2010}, we use the deconvolved source FWHM, $\ell_{\rm src}$, as the characteristic effective radius for the density calculation, rather than interpreting it as a diameter. This choice better matches the spatial extent over which the integrated flux is measured. Using $\ell_{\rm src}/2$ instead would mainly characterize the inner FWHM region and would overestimate the mean density. For the 209 resolved cores, the median number density is $5.2\times10^6$~cm$^{-3}$, with values ranging from $3.0\times10^4$ to $3.3\times10^8$~cm$^{-3}$.

The uncertainties of physical derivatives are quantified using Monte Carlo simulations. For each parameter, the input basic parameters are randomly sampled according to their measured uncertainties and the derived value was recalculated for each realization. The final uncertainty was obtained by adopting the standard deviation. We considered 18\% uncertainty for distance estimate, 24\% for gas-to-dust ratio estimate, 10\% for the environmental temperature estimate, 14\% for measured flux estimate. The physical derivatives and their uncertainties are listed in Table~\ref{tab:physics}. The systematic effect of unresolved core heating is assessed separately in Sect.~\ref{discuss:hierachy}.

\section{Mass transfer efficiency from cloud to core} \label{discuss:hierachy}

We characterize the distribution of mass across the cloud hierarchy using mass transfer efficiencies \citep{Motte2026}. For each subregion, we define the efficiency between two scales as the ratio of the total mass identified at the smaller scale to that at the larger scale. These are ratios of the observed masses at the time of observation. For the cloud ($\sim4$~pc) and dense-core ($\sim0.01$~pc) scales, we define
\begin{equation} \label{eq:eff_core}
\epsilon_{\rm core} = \frac{M_{\rm core}}{M_{\rm cloud}},
\end{equation}
where $M_{\rm core}$ is the summed mass of the dense cores in a subregion and $M_{\rm cloud}$ is its cloud mass. 

These efficiencies depend on the physical scales, angular resolution and source definitions used in the measurements \citep{ALMAGAL5}. The three G316.8 subregions provide a common framework for comparing them across environments at different evolutionary stages.

The cloud-to-core mass transfer efficiency $\epsilon_{\rm core}$ rises from \revise{$0.40\pm0.05\%$} to \revise{$1.54\pm0.07\%$} and $3.10\pm0.16\%$  from the young to intermediate and evolved subregions. The result aligns with the dynamic scenario of core growth as reported in recent ALMA large surveys (e.g., \citealp{Xu2024Assemble1} and \citealp{ALMAGAL5}) as well as in numerical simulations (e.g., \citealp{Chen2015}\revise{,} \citealp{Rosetta3}\revise{, and \citealp{Li2026}}). 

\begin{figure*}[!ht]
\centering
\includegraphics[width=0.33\linewidth]{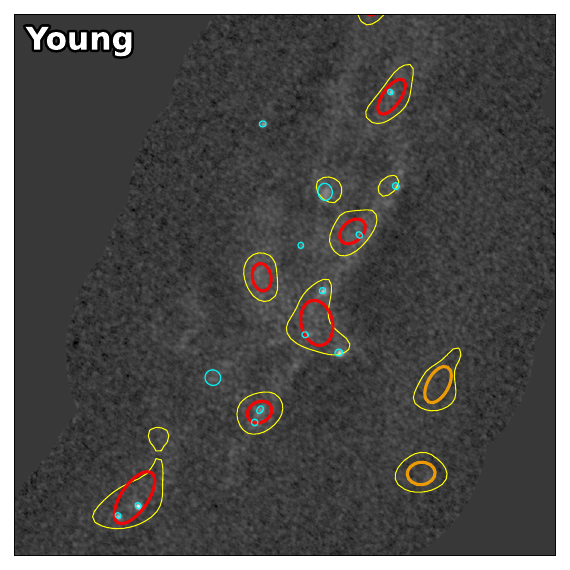}
\includegraphics[width=0.33\linewidth]{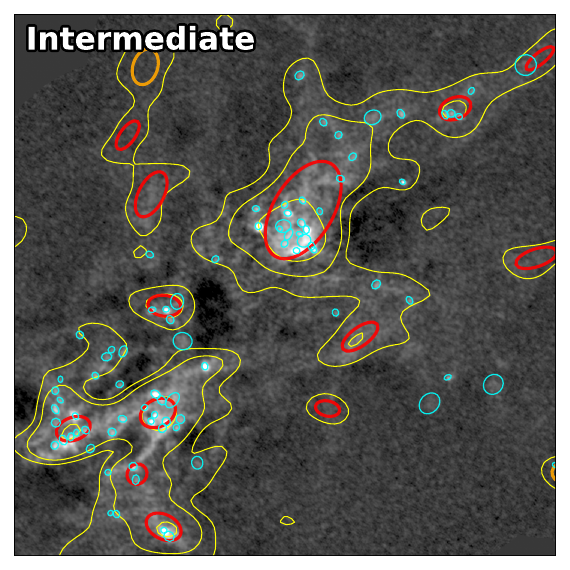}
\includegraphics[width=0.33\linewidth]{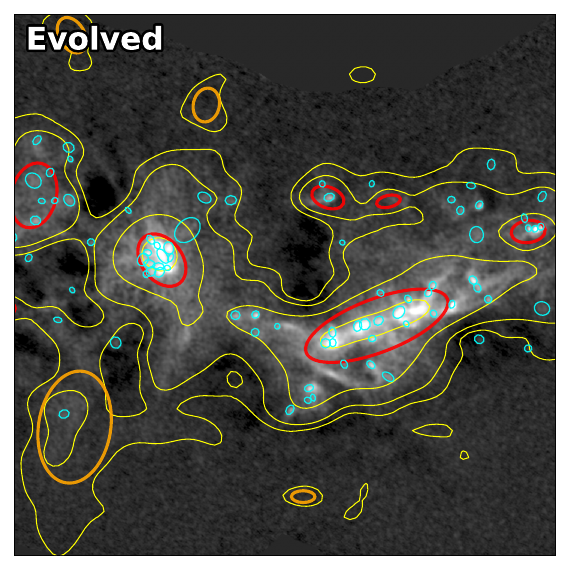}
\caption{{\bf Multi-scale structural identification in three evolutionary stages.} Background shows the {\tt 12m7m} 1.3 mm continuum emission. Yellow contours display the ACA 7~m continuum emission, with levels of 3, 12, 48, and 96 times $\sigma=2 \mjybeam$. Red ellipses show Paper~I fragments whose entire dendrogram leaf footprints lie within the 12~m+7~m primary-beam response of 0.2; orange ellipses show fragments extending beyond this boundary, which are excluded from the mass comparison. Cyan ellipses show the dense cores identified in this work.}
\label{fig:cloud2core}
\end{figure*}

Cloud as integral of total gas reservoir, can fragment into substructures but not necessarily immediately into cores. Those substructures, captured by recent lower-resolution interferometric studies (e.g., \citealp{Xu2024Quarks2} and \citealp{Morii2024}), are called (sub-cloud/clump) fragments. From their mass--radius relation, it was argued that these structures trace the physical scale at which gravity overtakes turbulence and a steep density profile develops. In this work, we retrieve the catalogue of fragments from Paper~I, which are identified in the ACA 7~m continuum image by \texttt{astrodendro} algorithm. We retain fragments whose entire dendrogram leaf footprints lie within the 12~m+7~m primary-beam response of 0.2: 10, 27, and 12 in the young, intermediate, and evolved subregions, respectively. The selected fragments are shown as red ellipses in Fig.~\ref{fig:cloud2core}. As a direct comparison, dense cores identified in this work are indicated by cyan ellipses. 

To estimate possible free--free bias in the core and fragment masses, we use the native $2.5\arcsec$ CORNISH-South 5.5~GHz map, extrapolated to 223.8~GHz assuming optically thin emission with $S_\nu\propto\nu^{-0.1}$. We assume smooth surface brightness below the radio beam to assign emission to the smaller core footprints, and match the template to the larger ACA beam for the fragments. With an adopted radio rms of 1.0~mJy~beam$^{-1}$ and a $3\sigma$ detection threshold, 32 of the 94 covered evolved-region cores overlap detected radio emission; the remaining 62 have nominal $1\sigma$ aperture limits. Ten of the twelve retained evolved-region fragments have complete radio coverage.

At fixed temperature and opacity, the inferred fragment mass changes from 1323 to 1191~$M_\odot$ (9.9\%). Applying the original-image aperture fractions to the adopted core masses as a sensitivity scenario changes the regional core mass from 248 to 240~$M_\odot$ (3.3\%). These estimates assess the sensitivity of the mass comparison to the assumed free--free distribution; the detailed procedure is given in Appendix~\ref{app:ff}.

Including fragments at an intermediate scale of $\sim0.15$~pc, we express the cloud-to-core efficiency as the product of the cloud-to-fragment and fragment-to-core mass transfer efficiencies:
\begin{equation} \label{eq:split_eff}
\begin{split}
\epsilon_{\rm core} 
= \epsilon_{\rm sub} \times \epsilon_{\rm com}
= \frac{M_{\rm sub}}{M_{\rm cloud}} \times \frac{M_{\rm core}}{M_{\rm sub}},
\end{split}
\end{equation}
Here $M_{\rm sub}$ is the summed mass of selected fragments in the subregion, $\epsilon_{\rm sub}=M_{\rm sub}/M_{\rm cloud}$, and $\epsilon_{\rm com}=M_{\rm core}/M_{\rm sub}$. Each ratio compares regional mass sums at the indicated scales.

\begin{figure}
\centering
\includegraphics[width=\linewidth]{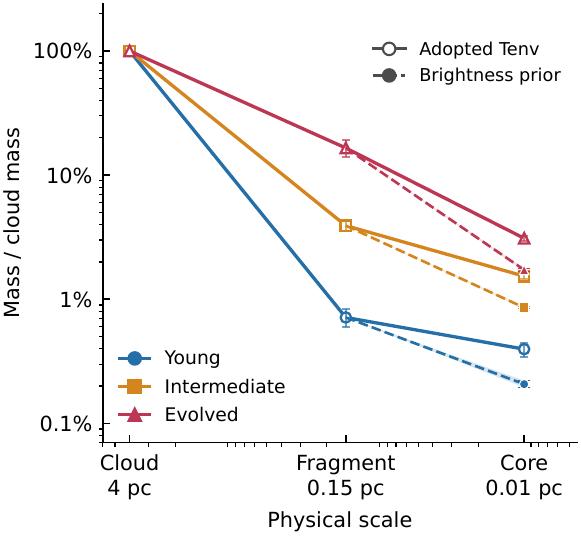}
\caption{{\bf Hierarchical mass transfer efficiency.} Mass ratios relative to the cloud-scale gas reservoir are shown as a function of physical scale for the three evolutionary subregions (young, intermediate, and evolved). Each curve connects the cloud-normalized masses at the cloud ($\sim4$~pc), fragment ($\sim0.15$~pc), and core ($\sim0.01$~pc) scales. Solid curves and open markers use $T_{\rm c}=T_{\rm env}$, with counting uncertainties shown by thin error bars. Dashed curves and filled core markers show the median fractions from the brightness-dependent temperature test with regional upper temperatures of 30, 40 and 50~K. Shading and the corresponding core error bars span the 16th--84th percentiles for independent core-temperature draws, holding fragment and cloud masses fixed. Curves and shading guide the eye between reservoir scales. Fragment masses include whole catalogue sources whose entire dendrogram leaf footprints lie within the 12~m+7~m primary-beam response of 0.2; the cloud masses refer to the full subregions.}
\label{multiscale_eff}
\end{figure}

Fig.~\ref{multiscale_eff} shows the mass ratios relative to the cloud reservoir at the cloud ($\sim4$~pc), fragment ($\sim0.15$~pc), and core ($\sim0.01$~pc) scales. These ratios decrease toward smaller scales in each subregion, showing how the mass is distributed across the hierarchy. 

Across the evolutionary sequence, the two mass transfer efficiencies exhibit opposite trends. The cloud-to-fragment efficiency $\epsilon_{\rm sub}$ rises from $0.7 \pm 0.1\%$ to $3.9 \pm 0.4\%$ and $16.5 \pm 2.5\%$ from the young to the intermediate and evolved regions. The fragment-to-core efficiency $\epsilon_{\rm com}$ decreases from \revise{$55.6 \pm 9.3\%$} to \revise{$39.2 \pm 3.2\%$} and $18.7 \pm 1.8\%$. Thus, the fraction of cloud mass identified in fragments increases along the sequence, while the ratio of core mass to fragment mass decreases. This behavior is illustrated in Fig.~\ref{fig:cloud2core}: the dense fragments enclosed by the yellow contours become increasingly extended, whereas the dense cores remain confined to localized peaks. The growing amount of extended emission included within fragments but outside dense cores is consistent with the declining fragment-to-core mass ratio. A qualitatively similar evolutionary behavior was reported by \citet{Dib2023} for the ALMA-IMF sample, where the characteristic size of overdense star-forming structures increases from $\sim0.05$ to $\sim0.5$~pc toward more evolved protocluster regions. In the case of M16 Eagle Nebula, \citet{Pervaiz2026} argued that dense structures like filaments form prior to widespread dense core formation as an intermediate step of structural formation.

The regional ordering in Fig.~\ref{multiscale_eff} is retained in a sensitivity test allowing for free--free contamination. Under the smooth-emission scenario above, the cloud-to-fragment efficiency changes from 16.5\% to 14.9\%, the cloud-to-core efficiency from 3.10\% to 3.00\%, and the fragment-to-core efficiency from 18.7\% to 20.1\%. Thus, accounting for contamination under this model preserves the opposing regional trends in the two mass fractions.

We quantify the temperature dependence of these mass fractions using a core-heating test motivated by the ALMA-IMF high-resolution temperature studies \citep{Dellova2024}. We assign $T_{{\rm c},i}=T_{{\rm env},i}+F_i(T_{{\rm upp},r}-T_{{\rm env},i})$, adopting upper temperatures of 30, 40 and 50~K for the young, intermediate and evolved regions. We compare uniform sampling, $F_i\sim U(0,1)$, with a brightness-dependent distribution, $F_i\sim{\rm Beta}(1+2q_i,3-2q_i)$, where $q_i$ is the peak-brightness percentile across the core catalogue. The latter assigns a mean heating fraction increasing from 25\% to 75\% with brightness. For each model, $10^5$ Monte Carlo realizations recalculate core masses.

The brightness-dependent model gives summed core masses of \revise{$22.8\pm1.4$, $145.6\pm3.5$ and $138.1\pm3.5$}~$M_\odot$ in the young, intermediate and evolved regions, respectively; the intervals span the 16th--84th percentiles of the temperature realizations. The corresponding median $\epsilon_{\rm core}$ values are \revise{0.21\%, 0.86\% and 1.73\%}, and $\epsilon_{\rm com}$ becomes \revise{29.0\%, 21.9\%} and 10.4\% (Fig.~\ref{multiscale_eff}). Uniform temperature sampling gives the same regional ordering of the median fractions. Thus core heating lowers the absolute core-mass fractions while preserving the opposing median trends across the three regions.

Shared regional heating increases the systematic spread. Assigning a common heating percentile to cores within each region gives 16th--84th percentile mass ranges of \revise{18.2--27.9, 119.8--173.4 and 116.8--160.7}~$M_\odot$ for the brightness-dependent model. The median trends persist, with greater overlap between regions and changes in their ordering in individual realizations. These ranges describe the sensitivity to the assumed heating distribution and its regional dependence.

We offer a possible explanation for the low mass transfer efficiency from fragment to core scales. Inspired by \citet{VS2024}, mass transfer toward smaller scales need not be fully efficient, because part of the accreted material may ``stagnate'' rather than being transported inward immediately, a process they refer to as gravitational choking. In our case, this intermediate reservoir would correspond to the fragment or intercore-medium scale. Indeed, from the continuity equation, any density profile $\rho(r) \propto r^{-p}$ with $p$ shallower than 2 can, in principle, lead to such inefficient inward transfer. Further progress will require higher-resolution temperature to test this interpretation more robustly. 

\section{Environmental regulation on core formation} \label{discuss:env}

Dense cores are integrated with their host environments by a number of intertwined physical quantities such as density, temperature, magnetic fields, and turbulent support. These global physics can determine local core growth, fragmentation, and subsequent collapse in theoretical models (e.g., \citealp{Bonnell2001}, \citealp{Padoan2002}, \citealp{McKee2003}, and \citealp{VS2019}). Such physical interpretation could lead to observable correlations between the properties of dense cores and molecular clouds. Identifying these core--environment correlations is thus a promising avenue for understanding the physics governing dense core evolution and, consequently, star formation and cloud evolution.

To characterize the environment around each core, we use a circular aperture centered on its position, with a diameter of $30\arcsec$ ($0.4$ pc). This size is comparable to the resolution of the APEX C$^{18}$O~(2--1) velocity-dispersion map, the coarsest environmental tracer used here. For each physical-quantity map, the aperture median is the median of the positive, finite pixel values within the aperture; it describes the typical mapped value around the core. The scaled median absolute deviation measures spatial variation within the aperture. The environmental quantities used in this analysis are:
\begin{enumerate}
\item ${T}_{\rm env}$: Aperture-median dust temperature from the far-infrared SED fitting in Paper~I. The core mass calculation in Sect.~\ref{cat:property} instead uses the temperature sampled at the core peak.
\item ${\sigma}_{\rm env}$: Aperture median of the line-of-sight velocity dispersion derived from the APEX C$^{18}$O $(2-1)$ observations (Appendix~\ref{app:sigma}).
\item ${\Sigma}_{\rm env}$: Aperture-median gas surface density from SED fitting.
\item ${\Sigma}^{\rm ic}_{\rm env}$: Aperture median of the intercore surface-density map, obtained by converting the source-removed $I^{\rm ic}_{\nu}$ map using Eq.~\ref{eq:surfden}.
\end{enumerate}
Core-scale physical parameters used in this analysis include:
\begin{enumerate}
\item ${M}_{\rm core}$: Core mass.
\item ${\Sigma}_{\rm core}$: Core surface density.
\item ${n}_{\rm core}$: Core volume-averaged H$_2$ number density.
\item ${\Sigma}^{\rm ic}_{\rm core}$: Local intercore surface density at the core peak (Sect.~\ref{cat:property}), sampling the source-removed map at its $0.71\arcsec\times0.61\arcsec$ beam resolution ($\sim0.009$~pc), whereas ${\Sigma}^{\rm ic}_{\rm env}$ is the median of that map within a $30\arcsec$-diameter (0.4~pc) aperture.
\end{enumerate}

\begin{figure}
\centering
\includegraphics[width=\linewidth]{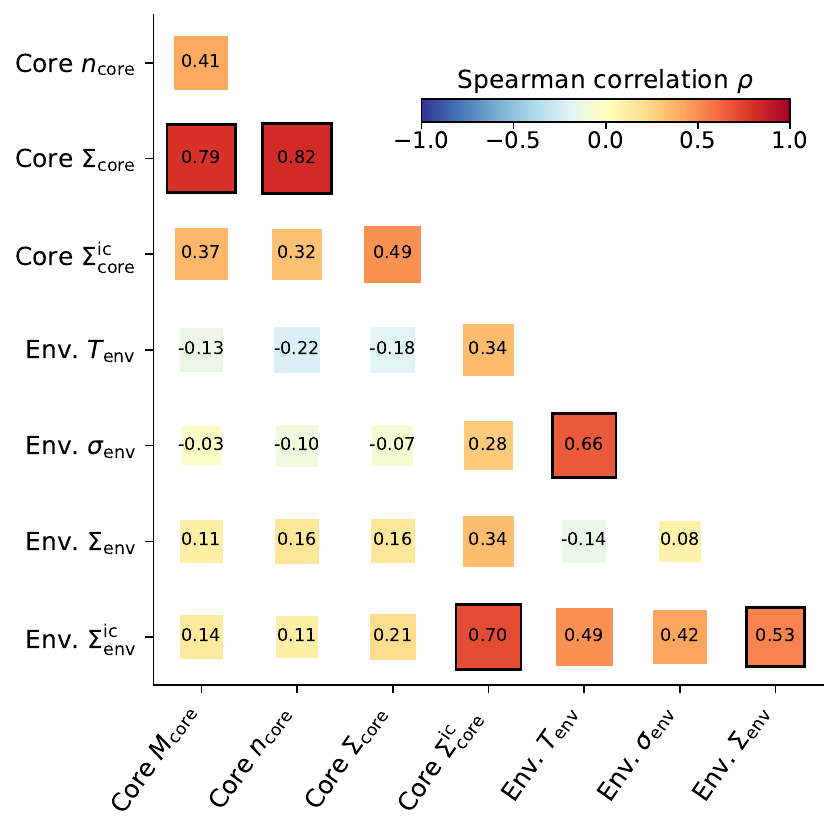}
\caption{{\bf Correlation matrix of core and environment physics.} Each square represents one parameter pair in log-log scale, with the color showing the value of the Spearman coefficient and the square size scaled by its absolute magnitude. Pairs with absolute coefficients above \revise{0.5} are marked with black outlines.}
\label{fig:corner}
\end{figure}

Fig.~\ref{fig:corner} shows correlations between every parameter pair. Pairs involving $n_{\rm core}$ use the resolved sample (209 out of 231), with pairwise filtering of non-positive or invalid values; surface-density, mass and environmental quantities retain all valid measurements. At the core scale, the core surface density $\Sigma_{\rm core}$ is strongly correlated with both core mass $ M_{\rm core}$ and core volume density $ n_{\rm core}$, with coefficients of 0.79 and \revise{0.82}, respectively. The direct correlation between mass and volume density is weaker, with a coefficient of \revise{0.41}. Since dense core sizes are similar, the mass dependence of the three parameters by definition should result in correlation. In addition, the core surface density shows a moderate positive correlation with the core local medium surface density, $\Sigma_{\rm core}^{\rm ic}$, with a coefficient of \revise{0.49}. Actually, this correlation should be a natural outcome of the continuity of ISM, where denser and more massive cores tend to emerge in high-density regions.

Among the environmental parameters, the intercore surface density, $\Sigma_{\rm env}^{\rm ic}$, is the quantity that shows the broadest correlation with others. It exhibits moderate positive correlations, all with coefficients close to 0.5, with the environmental dust temperature, $ T_{\rm env}$, the velocity dispersion, $\sigma_{\rm env}$, and the total environmental surface density, $\Sigma_{\rm env}$. These relations indicate that the intercore medium relies on the surrounding environment determined by larger-scale cloud properties.

\begin{figure}[!h]
\centering
\includegraphics[width=\linewidth]{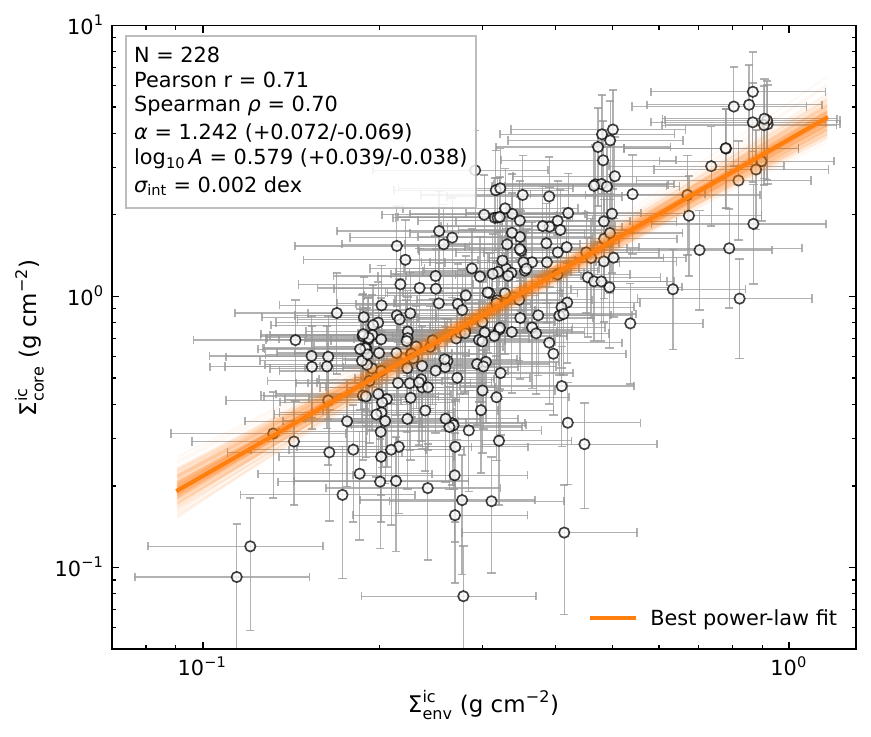}
\caption{{\bf Correlation between $\Sigma_{\rm env}^{\rm ic}$ and $\Sigma_{\rm core}^{\rm ic}$}. The hollow-circle data points represent 228 cores valid used for correlation, all with error bars giving data uncertainties. Here $\Sigma_{\rm env}^{\rm ic}$ is the median of the intercore surface-density map within a $30\arcsec$-diameter aperture, and $\Sigma_{\rm core}^{\rm ic}$ samples the same map at the core peak with a $0.71\arcsec\times0.61\arcsec$ beam ($\sim0.009$~pc). The aperture diameter corresponds to 0.4~pc. The orange solid line marks the best fitting power-law relation obtained from the MCMC fitting, while low-opacity curves show a set of posterior realizations giving the allowed range of fitted relations. The number of valid data points for fitting, Pearson and Spearman correlation coefficients, slope, intercept and intrinsic scatter are shown in the top left corner.}
\label{fig:corr_env_core}
\end{figure}

The cross-scale correlations to link cores and environs are the $4\times4$ grid in the bottom left corner of Fig.~\ref{fig:corner}. The most significant correlation is the one between the environmental intercore surface density, $\Sigma_{\rm env}^{\rm ic}$, and the core immediate surrounding material surface density, $\Sigma_{\rm core}^{\rm ic}$, with a correlation coefficient of \revise{0.70}. These two estimators sample the same intercore map on nested scales, so a positive correlation is expected. The relation quantifies how the local intercore emission at core positions is enhanced relative to the typical emission within the surrounding 0.4~pc aperture. To quantify this relation, we fitted a power-law model using a Markov chain Monte Carlo (MCMC) approach. The fit is performed in logarithmic space, where a power-law relation can be represented as a linear one, and the measurement uncertainties in both variables are propagated and incorporated into the likelihood evaluation. Fig.~\ref{fig:corr_env_core} shows the resulting best-fitting relation together with its uncertainty range,
\begin{equation}
\log_{10}\Sigma_{\rm core}^{\rm ic} = \revise{1.242^{+0.072}_{-0.069}}\times \log_{10}\Sigma_{\rm env}^{\rm ic} + \revise{0.579^{+0.039}_{-0.038}}.
\end{equation}
The $p$-values of both Pearson and Spearman are of order \revise{$10^{-35}$}$\ll0.05$, showing statistically significant correlations. The inferred intrinsic scatter, $\sigma_{\rm int}$, is very small, indicating that the observed dispersion can already account for by the measurement uncertainties alone, without requiring any additional intrinsic scatter. Moreover, the best-fitting slope is slightly steeper than 1, suggesting that $\Sigma_{\rm env}^{\rm ic}$ and $\Sigma_{\rm core}^{\rm ic}$ are super-linearly correlated. As a reference, at $\Sigma_{\rm env}^{\rm ic}=1$~\gpsc, the fitted relation gives $\Sigma_{\rm core}^{\rm ic}\simeq3.8$~\gpsc.

By contrast, the direct correlations between the other environmental parameters and the intrinsic core properties are generally weak or absent. The strongest cross-scale relation therefore describes the spatial structure of the intercore medium itself. Together with the correlation between compact-core and local intercore surface densities, it is consistent with a connection between core formation and the surrounding gas concentration.

The correlations discussed above suggest the following empirical links between the environment and core properties:
\[
\underbrace{\left( T_{\rm env},\, \sigma_{\rm env},\, \Sigma_{\rm env}\right)
\;\rightarrow\;
\Sigma_{\rm env}^{\rm ic}}_{\mathrm{Environment\;physics}}
\;\rightarrow\;
\underbrace{\Sigma_{\rm core}^{\rm ic}
\;\rightarrow\;
\left(\Sigma_{\rm core},\,  M_{\rm core},\,  n_{\rm core}\right)}_{{\rm Core\; physics}}
\]
In this scenario, the core-forming environment, characterized by ${T}_{\rm env}$, $\sigma_{\rm env}$, and $\Sigma_{\rm env}$, is moderately connected to the intercore medium at sub-pc scale through $\Sigma_{\rm env}^{\rm ic}$. The intercore medium then serves as a bridge to the core-immediate surrounding gas $\Sigma_{\rm core}^{\rm ic}$. At the core scale, $\Sigma_{\rm core}^{\rm ic}$ is further linked to core properties such as the surface density $\Sigma_{\rm core}$ and mass $ M_{\rm core}$. 

The physical scales of the intercore medium are comparable to, or slightly smaller than, those of clumps, which are commonly regarded as the pc-scale reservoirs of star-forming gas \citep{Urquhart2022}. In this sense, the env--core correlation found here can also be viewed as a clump--core correlation, similar to those reported by \citealt{Traficante2023} and \citealt{ALMAGAL5}. A key difference, however, is that those studies focused primarily on the most massive core within each reservoir, since all cores in a given region were associated with the same parental clump. In our case, by contrast, each core is treated individually, with its local environment defined in a consistent way thanks to the large mosaic coverage and high-fidelity image combination. Our data therefore provide a more general view of environment-regulated core formation. We note, however, that the arrows in the schematic picture represent empirical links inferred from the correlation analysis rather than direct proof of causality. Establishing the physical connection across scales will require future observations of multi-scale gas inflow to trace how material is transported from the reservoir down to the core scale (e.g. \citealp{Xu2023SDC335, Beuther2025b}).

\section{Caveats}

Several caveats should be considered. First, the $2.5\arcsec$ CORNISH-South beam does not resolve the dense-core scale. The assumed smooth distribution of free--free surface brightness below this beam cannot be tested with these data, so the inferred contribution within individual core footprints depends on that assumption. Radio observations at a resolution comparable to the ALMA data are needed to constrain the sub-beam distribution directly. Second, the Herschel-based temperature map measures the core environment. The heating distributions used in Sect.~\ref{discuss:hierachy} parameterize unresolved core temperatures, while higher-resolution temperature tracers are needed to calibrate these distributions and their regional dependence. Finally, the physical nature of the fragments, dense cores, and intercore medium remains to be tested dynamically. Their mass--radius relation suggests that many fragments are likely gravity-dominated structures, but molecular-line kinematics are needed to confirm whether they are bound, collapsing, or transient. In particular, the intercore medium may represent more than a decomposition-defined background component: as the material linking pc-scale cloud structure to compact dense cores, it may mark a dynamically important scale at which gas becomes partially decoupled from the larger cloud and more directly connected to core-scale mass assembly.

\section{Conclusion} \label{sec:conclude}

The G316.8 filament provides an evolutionary sequence of massive star and cluster formation, from a northern infrared-dark (young) region, through a high-mass young stellar object (intermediate) region, to a southern H{\sc ii} (evolved) region. Linear filament and nested cluster evolution tomography (LANCET) aims at leveraging this in-situ evolutionary sequence as a controlled set to understand dynamic processes of massive star formation. As LANCET Paper~II, we have presented a 1.3 mm continuum image of the massive filament G316.8, constructed by combining ALMA 12~m, ACA 7~m, and TP observations. By recovering emission from the 10-pc filamentary cloud down to thousand-au dense-core scales, the data enable a systematic investigation of hierarchical mass assembly and the environmental regulation of core formation. Our main findings are summarized as follows:

\begin{enumerate}
\item We extracted 231 sources from their background emission. Their spatial distribution varies systematically along the filamentary evolutionary sequence: the young region contains sparse and relatively isolated sources while evolved regions are showing more clustered configuration. Based on mass and sizes, these sources are classified as dense cores. Their statistics like mass function, spatial separation and mass segregation will be presented in the forthcoming papers. Source extraction also produced two background maps. The TP-containing background includes the large-scale emission surrounding the cores and supplies the foreground/background term in their optical-depth calculation. The \texttt{12m7m} background traces the intercore medium on local ($<0.4$~pc) scales. Together, they connect the compact sources to their surrounding filamentary cloud. 

\item We use mass transfer efficiencies, defined as ratios of the observed regional masses at different physical scales, to characterize the hierarchy from clouds ($\sim4$~pc) through fragments ($\sim0.15$~pc) to dense cores ($\sim0.01$~pc). From the young to the evolved region, the cloud-to-core efficiency rises from $0.4\%$ to $3.1\%$. The cloud-to-fragment efficiency increases from $0.7\%$ to $16.5\%$, while the fragment-to-core efficiency decreases from \revise{$55.6\%$} to $18.7\%$. These trends show an increasing fraction of cloud mass in fragments and a decreasing ratio of core mass to fragment mass along the evolutionary sequence.

\item For each core, we characterized the surrounding environment using aperture medians of dust temperature, total surface density, velocity dispersion, and intercore surface density within a 0.4 pc diameter. Among these environmental properties, the intercore surface density shows the broadest set of correlations with other quantities. Dust temperature and velocity dispersion show positive correlation possibly reflecting the influence of stellar feedback. The beam-averaged core surface density correlates strongly with mass across the full sample and with volume density in the resolved sample; the mass--volume-density correlation is weaker. The core surface density also correlates with the local intercore surface density at the core peak, with a coefficient of \revise{0.49}.

\item Among all environment--core pairs, the strongest correlation, with a coefficient of \revise{0.70}, is between the aperture-median intercore surface density and the intercore surface density at the core peak. A power-law fit gives a slope of \revise{$1.24\pm0.07$}, describing the concentration of the same intercore emission from the 0.4~pc aperture scale toward core positions sampled with the $\sim0.009$~pc beam. The remaining  pairs show no strong correlation. This relation measures the spatial concentration of the intercore medium. Together with the core--intercore surface-density correlation, it supports a link between dense cores and their surrounding gas distribution. 
\end{enumerate}

\section*{Data availability}
\revise{ALMA data are available at }\url{https://almascience.eso.org//aq/} by setting the observation code 2016.1.00909.S. The reduced ALMA data used for this study and generated Tables will be available \revise{in CDS} once the paper is published. The ancillary data are publicly accessible through the ESO archive for VVV (\url{https://doi.eso.org/10.18727/archive/67}), the Spitzer Heritage Archive (\url{https://irsa.ipac.caltech.edu/applications/Spitzer/SHA/}), the Herschel Science Archive via the IRSA search interface (\url{https://irsa.ipac.caltech.edu/applications/Herschel/}), and the SARAO MeerKAT Galactic Plane Survey data release (\url{https://doi.org/10.48479/3wfd-e270}).

\begin{acknowledgements}

\revise{The authors thank the anonymous referee for improving and polishing the article.} 
We thank Markus Nielbock, \revise{today} at the Haus der Astronomie, who worked for SEST/SIMBA 25 years ago for providing us with the spectral response of SIMBA, as measured in the lab. 
\revise{F.X. thanks Siyi Feng for providing discussion opportunities with her group member and therefore inspiring this project at Xiamen University in Nov. 2023.} 
This work has been supported by the National SKA Program of China (2025SKA0140100).
This work was supported by the CRC 1601 (sub-project B2) funded by the DFG -- 500700252 and the Cluster of Excellence ``Our Dynamic Universe'' (DYNAVERSE). 
G.G. gratefully acknowledges support by the ANID BASAL project FB210003.
Z.-Y.Z. acknowledges the support of the National Natural Science Foundation of China (NSFC) under grants No. 12533003,1257030642. Z.-Y.Z. is supported by the Fundamental Research Funds for the Central Universities (Grant no. KG202502).
This paper makes use of the following ALMA data: ADS/JAO.ALMA\#2016.1.00909.S. ALMA is a partnership of ESO (representing its member states), NSF (USA) and NINS (Japan), together with NRC (Canada), NSTC and ASIAA (Taiwan), and KASI (Republic of Korea), in cooperation with the Republic of Chile. The Joint ALMA Observatory is operated by ESO, AUI/NRAO and NAOJ. The National Radio Astronomy Observatory is a facility of the National Science Foundation operated under cooperative agreement by Associated Universities, Inc.
This work made use of Astropy: a community-developed core Python package and an ecosystem of tools and resources for astronomy \citep{Astropy2022}, Scipy \citep{Scipy2020}, Numpy \citep{Numpy2020}, emcee \citep{Foreman2013}. 

\end{acknowledgements}

\bibliographystyle{aa}
\bibliography{lancet}

\begin{appendix}

\section{12~m+7~m interferometric combination} \label{app:psf}

\begin{figure}[!ht]
\centering
\includegraphics[width=\linewidth]{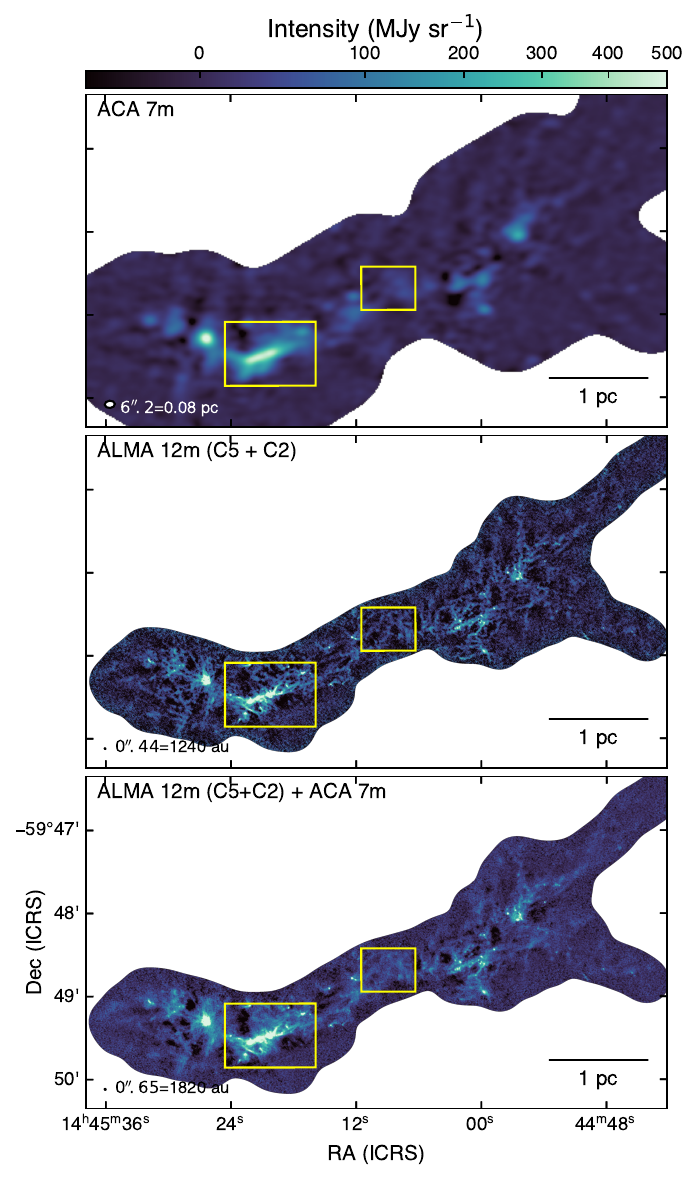}
\caption{{\bf Continuum images from different combinations of interferometric data.} {\it Top}--{\it Bottom}: ACA 7~m-only, ALMA 12~m-only, and 12~m + 7~m image. All the images are after the PB correction to show real sky brightness.}
\label{fig:interf_combine}
\end{figure}

Fig.~\ref{fig:interf_combine} compares the ACA 7~m-only, ALMA 12~m-only and combined images. The 7~m-only image recovers the overall filament skeleton and compact sources, but with insufficient resolution of $6\arcsec$ to resolve smaller structures. In contrast, the 12~m-only image, with more than an order-of-magnitude higher angular resolution, \revise{reveals dense cores and narrow, thousand-au-scale filamentary features}; however, it also exhibits discontinuous emission and negative artifacts caused by missing short spacings. 

The combined \texttt{12m7m} image resolves fine substructure along with the extended emission. This improvement is illustrated in the two regions highlighted by yellow rectangles: the left region shows extended emission perpendicular to a linear filament that is recovered in the \texttt{12m7m} image, while the right region shows faint web-like structures whose negative artifacts are largely removed after combination.

\begin{figure}[!h]
\centering
\includegraphics[width=\linewidth]{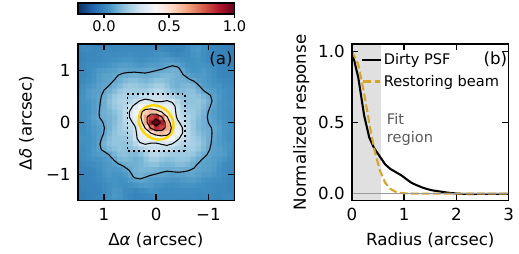}
\caption{{\bf The combined PSF of the 12~m+7~m imaging.} (a) Peak-normalized mosaic PSF within $\pm1.5\arcsec$. Solid contours are at 0.1, 0.3, 0.5, 0.7 and 0.9. The yellow ellipse shows the adopted restoring beam, and the dotted square marks the central $11\times11$-pixel fitting region. (b) Azimuthal profiles of the PSF (solid) and restoring Gaussian (dashed). The shaded band marks the half-width of the fitting region.}
\label{fig:psf}
\end{figure}

Fig.~\ref{fig:psf}(a) shows the final combined point-spread function (PSF), reconstructed with the adopted MIRIAD imaging parameters: robust weighting 0.5, a $0.1\arcsec$ Gaussian taper and $0.1\arcsec$ pixels. The effective mosaic PSF at the reference position was calculated with \texttt{mospsf}. Its narrow central peak is surrounded by a broader shoulder. At a radius of $0.75\arcsec$, where the azimuthally averaged restoring Gaussian has fallen below 5\% of its peak (approximately 0.03), the corresponding PSF amplitude is still 0.17. This comparison quantifies the excess response in the PSF wing. The peak-normalized 12~m-only PSF already reaches 0.14 at the same radius, showing that much of this shoulder is present before adding the synthetic 7~m visibilities. The combined profile first crosses zero at approximately $2\arcsec$; beyond this radius, the sidelobes remain within approximately 5\% of the peak. MIRIAD fits the restoring Gaussian to the central $11\times11$ pixels of the dirty beam, giving it as $0.71\arcsec\times0.61\arcsec$ at position angle \revise{$47.4^\circ$}. The azimuthally averaged profile of the restoring Gaussian is also shown in Fig.~\ref{fig:psf}(b) for comparison.

\section{Free-free 1.3 mm emission} \label{app:ff}

We constructed free-free emission model using the SARAO MeerKAT 1.3~GHz Galactic Plane Survey \citep[MGPS;][]{Goedhart2024} and the Australia Telescope Compact Array (ATCA) CORNISH-South survey \citep{Irabor2023}. The MGPS data provide 14 subbands (two are flagged due to Radio Frequency Interference) between 0.9 and 1.6 GHz, while the CORNISH-South data add a higher-frequency measurement at 5.5 GHz. MeerKAT has extensive short baseline coverage allowing the imaging of extended emission, so angular scales up to $10\arcmin$ are generally well recovered \citep{Goedhart2024}.

All maps were convolved to a common angular resolution of $8\arcsec$ and reprojected onto a common grid. For each pixel, subband flux densities were fitted with the free--free emission model given by, 
\begin{equation} \label{eq:S_nu}
    S_\nu = \Omega B_\nu (T_{\rm e}) \left(1 - e^{-\tau_{\rm ff}}\right),
\end{equation}
where $\Omega$ is the beam solid angle, $T_{\rm e}$ is the electron temperature, and $B_\nu (T_{\rm e})$ is the Planck function. In the radio regime, the Rayleigh-Jeans approximation is valid, such that $B_\nu (T_{\rm e}) \approx 2kT_{\rm e}\nu^2/c^2$. The optical depth is written as,
\begin{equation} \label{eq:tau_ff}
    \tau_{\rm ff} = 0.08235 \left(\frac{T_{\rm e}}{\mathrm{K}}\right)^{-1.35} \left(\frac{\nu}{\mathrm{GHz}}\right)^{-2.1} \left(\frac{EM}{\mathrm{pc\,cm^{-6}}}\right),
\end{equation}
where $EM=\int n_e^2 dl$ is the emission measure.

\begin{figure}[!h]
\centering
\includegraphics[width=0.8\linewidth]{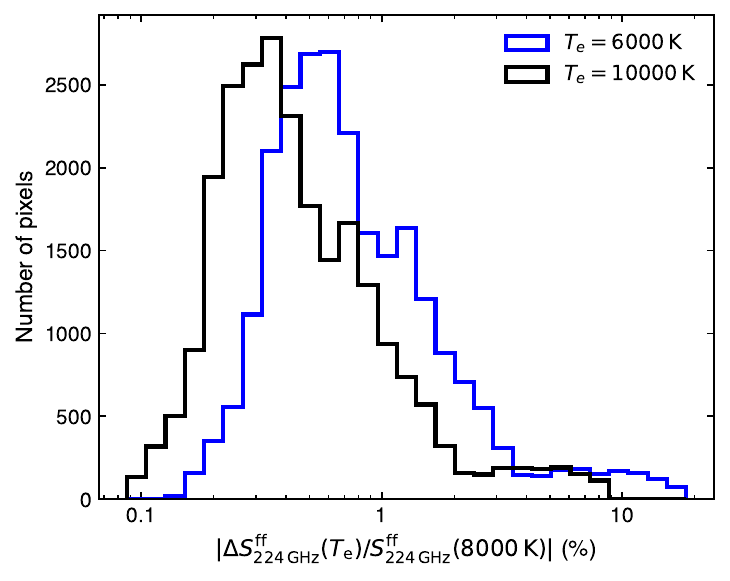}
\caption{{\bf Relative difference in free--free emission}. For each pixel, free-free emission is fitted under the electron temperature ($T_{\rm e}$) assumption of $6{,}000$~K, $8{,}000$~K and $10{,}000$~K, respectively. \revise{Histograms show the number of pixels against the absolute relative difference of the 223.8~GHz free--free emission to the fiducial case ($8{,}000$~K).}}
\label{fig:fftemplate_varyTe}
\end{figure}

Given the limited frequency coverage, $T_{\rm e}$ cannot be reliably constrained from the radio SED alone. We therefore adopted the empirical value with uncertainty $8{,}000\pm2{,}000$~K, consistent with typical Galactic H{\sc ii} regions \citep[e.g.][]{Garay1983, Quireza2006}, and fitted only $EM$. For each pixel, only subband measurements with signal-to-noise ratios greater than 5 were included in the fitting. The ATCA data suffer from significant missing flux due to coarse $uv$-coverage and were therefore used only in the central region. This is sufficient for our purposes, as the turn-over frequency is most likely to fall outside the MGPS frequency range only in these bright, compact regions, where the ATCA data provide essential constraints. The free--free emission model was then extrapolated to 1.3 mm. This pixel-by-pixel extrapolation yields a 1.3 mm free--free emission map, which is given on the left panel of Fig.~\ref{fig:synthetic}.

To assess the impact of assumed $T_{\rm e}$, we repeated the fitting procedure using $6{,}000$~K and $10{,}000$~K. The pixelwise difference in flux densities is plotted in Fig.~\ref{fig:fftemplate_varyTe}. \revise{The median differences are $0.7\%$ ($6{,}000$~K) and $0.4\%$ ($10{,}000$~K). All pixels lie within $10\%$ for $10{,}000$~K, and $98\%$ for $6{,}000$~K, with the remaining $1.7\%$ at $10{-}20\%$,} indicating that the inferred millimeter free--free contribution is only weakly sensitive to the assumed $T_{\rm e}$. Therefore, the uncertainty of the 1.3 mm map (Fig.~\ref{fig:synthetic}) is $\lesssim10\%$.

For the mass-contamination estimates in Sect.~\ref{discuss:hierachy}, we use the native $2.5\arcsec$ CORNISH-South 5.5~GHz map separately from the $8\arcsec$ SED fit. We adopt $\sigma=1.0$~mJy~beam$^{-1}$ and retain native radio pixels above $3\sigma$ and extrapolate their emission to 223.8~GHz with spectral index $-0.1$. Assuming smooth surface brightness below $2.5\arcsec$, we interpolate onto the ALMA pixel grid and integrate within each dense-core footprint using the corresponding solid angle. The angular resolution remains $2.5\arcsec$ after interpolation. The comparison flux is integrated directly from the original \texttt{12m7m} image over the same footprint, including its background. For covered radio nondetections, the nominal $1\sigma$ surface-brightness limit is converted to an aperture-integrated limit under the same smooth-emission assumption.

For the ACA fragments, we convolve the thresholded radio template to the $7.38\arcsec\times5.28\arcsec$ ACA beam and integrate within the original dendrogram leaf masks, applying the ACA primary beam to match the catalogue flux convention. The fractional contributions are converted to fragment-mass changes at fixed temperature and opacity. For the core-mass sensitivity scenario, each adopted core mass is scaled by one minus its original-image aperture fraction, holding its temperature, geometry and optical-depth factor fixed. Sources without radio coverage are held unchanged. This scenario does not directly recalibrate the background-subtracted core photometry.

\section{Total-power reconstruction} \label{app:TP}

For the G316.8 region, the only available TP continuum observation at this wavelength is the one obtained with the SEST Imaging Bolometer Array \citep[SIMBA;][]{Nyman2001}. However, two caveats of SIMBA data exist: One is its negative bowls around the source emission, which is due to atmospheric removal; the other is its broad bandpass, which may include CO~(2--1) line contamination at 230~GHz. In this work, we therefore construct a new 1.3 mm continuum model \revise{at the ALMA continuum reference frequency of 223.8~GHz,} with an angular resolution of $18\arcsec$\revise{,} by combining dust and free--free emission components. The dust emission component is extrapolated from far-infrared SED, assuming one dust thermal emission component (in Paper~I). To quantify the contribution of ionized gas, we reconstruct the free--free emission SED using radio continuum data as explained in Appendix~\ref{app:ff}. \revise{Beyond the coverage of the Paper~I dust model, the model is continued with dust emission from the 18\arcsec\ and 36\arcsec\ SED maps, at 36\arcsec\ resolution, so that it fills the 12m+7m mosaic and its surroundings.} 

\begin{figure}[!h]
\centering
\includegraphics[width=\linewidth]{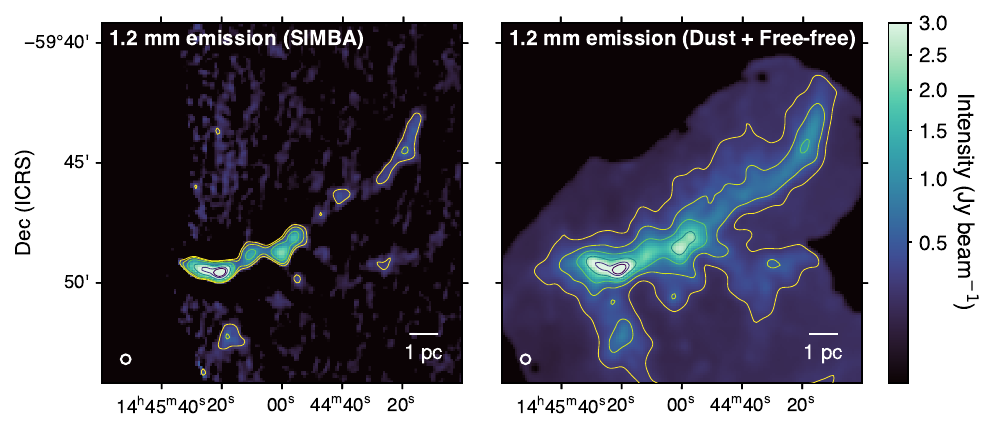}\\
\includegraphics[width=0.75\linewidth]{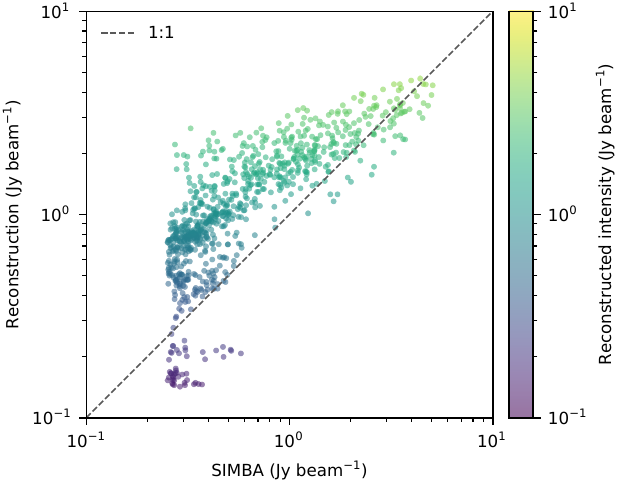}
\caption{\textbf{Zero-spacing information benchmark.} Top: SIMBA (left) and reconstructed (right) 1.2 mm continuum images at $24\arcsec$ resolution. The square-root color scale spans 0.05 to 3.0~\jybeam, and contours are at 0.25, 0.5, 1.0, 2.0, 3.0, and 3.5~\jybeam. Bottom: pixel-by-pixel comparison over the displayed field for positions above the SIMBA contour levels. The color indicates the reconstructed intensity, and the dashed line indicates equality.}
\label{fig:benchmark}
\end{figure}

\begin{figure*}[!ht]
\centering
\includegraphics[width=\linewidth]{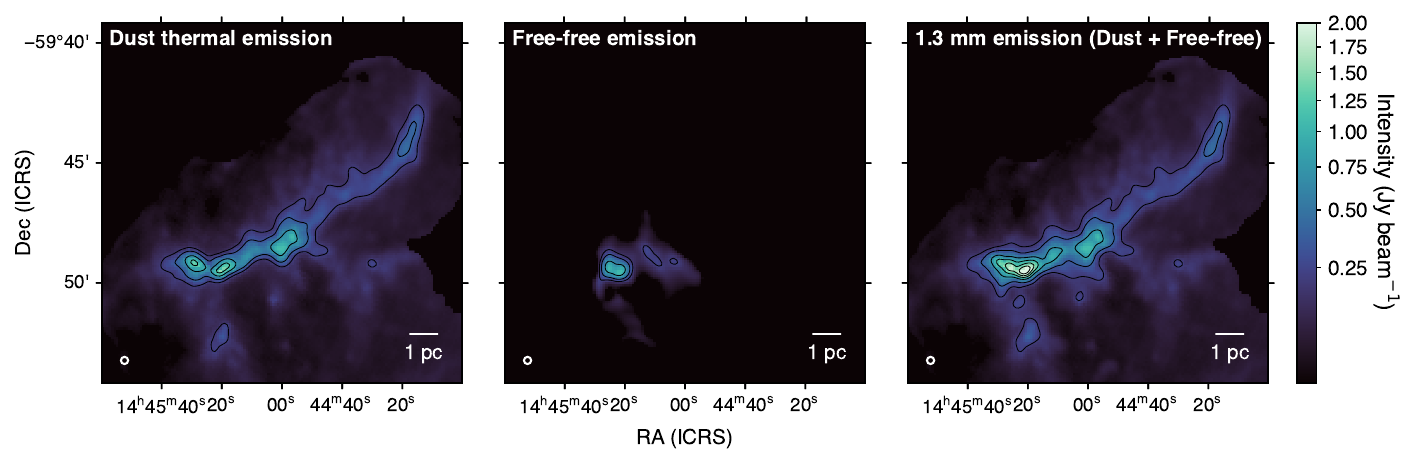}
\caption{{\bf Comparison of 1.3 mm continuum emission maps.} {\it Left}: Dust thermal emission at 1.3 mm extrapolated from far-infrared SED. {\it Middle}: Free-free emission at 1.3 mm extrapolated from centimeter SED. {\it Right}: Reconstructed 1.3 mm continuum emission by adding dust thermal and free-free emission components. All maps have a common angular resolution of $18\arcsec$, as indicated by white ellipses. The color scale is identical across panels. Black contours trace the emission distribution in each panel, with levels of 0.2, 0.35, 0.70, 1.05, 1.40, and 1.75 \revise{\jybeam}. The beam and scale bar of 1 pc are labeled in the bottom.}
\label{fig:synthetic}
\end{figure*}

As shown in Fig.~\ref{fig:synthetic}, 1.3 mm dust thermal emission dominates the large-scale filamentary structure, while free--free emission primarily contributes to the southern \ion{H}{ii} region. By adding two, we obtain the 1.3 mm continuum map. Besides, the inclusion of the free--free component changes the morphology in the southern region, where the emission changes from a double-peaked structure in the pure dust map to a single dominant peak in the combined map. The brightest peak is at RA(2000)= 14$^h$45$^m$20.82$^s$, DEC(2000)=-59$^\circ$49$'$28.77$''$ (ICRS), again consistent with SIMBA real observation. This agreement provides a validation of considering both dust and free--free emission in our case.

We benchmark the brightness of the TP reconstruction using the archival SIMBA observations toward G316.8. The observations were carried out on 28 October 2001, with Uranus used as the flux calibrator. The data reduction was performed with the MOPSI package developed by Robert Zylka (IRAM, Grenoble). The maps were reduced by applying atmospheric opacity corrections, fitting and subtracting baselines, and removing correlated sky noise. We followed the three-stage procedure recommended in the SIMBA manual. In the first iteration, all data were used for sky-noise removal. In the second iteration, regions showing source emission in the interim map were masked during the sky-noise removal. From this second interim map, a source model was derived and then included in the third iteration. The representative continuum frequency is 250~GHz, and the beam FWHM is 24\arcsec. The final 1.2 mm SIMBA image is shown in the left panel of Fig.~\ref{fig:benchmark}.

We used the same model to produce an expected 1.2 mm continuum map and then smoothed it to the SIMBA resolution of 24\arcsec, as shown in the upper right panel of Fig.~\ref{fig:benchmark}. The lower panel compares the brightness of corresponding pixels above the 0.25~\jybeam\ SIMBA threshold.

\section{Source extraction} \label{app:getsf}

We performed a \texttt{getsf} test in a $50\arcsec \times 50\arcsec$ sized box, but for three different images: (i) 12~m-only data (hereafter {\tt 12m}), (ii) 12~m and 7~m combined data ({\tt 12m7m}), and (iii) 12~m, 7~m, and TP combined data ({\tt 12m7mTP}). We followed the workflow described in \citet{Duet1}, and combined the three catalogs for comparison. Excluding sources whose peaks lie within $0.5\arcsec$ of the test-field boundary, the \texttt{12m}, \texttt{12m7m}, and \texttt{12m7mTP} catalogues contain 18, 25, and 27 sources, respectively.

\begin{figure}[!h]
\centering
\includegraphics[width=\linewidth]{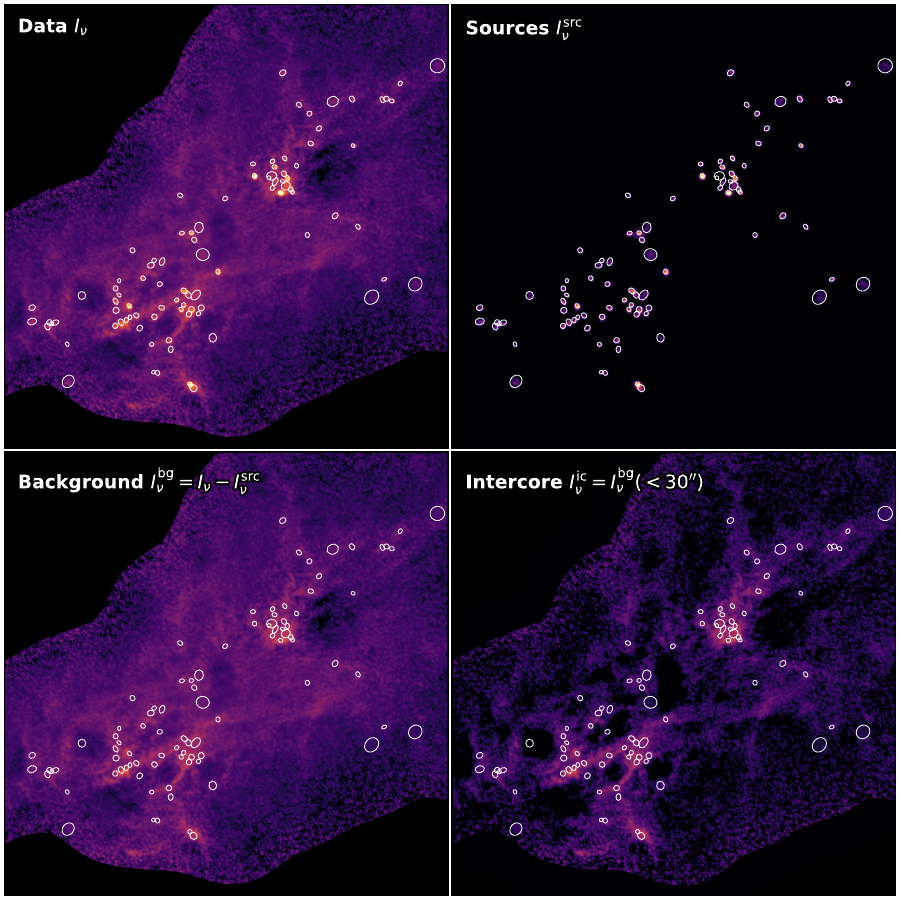}
\caption{{\bf Decomposition of source and background.} The \texttt{12m7mTP} observation data (top left) is decomposed into source (top right) and background (bottom left) components using the \texttt{getsf} algorithm. \revise{The panels show the intermediate $2\arcmin\times2\arcmin$ subregion on the colour scale of Fig.~\ref{fig:allscale}; the source and intercore components are divided by the 12m+7m mosaic gain.} For \texttt{12m7m} data, source component is the same but the background component is rather intercore background emission (bottom right). Detected sources are highlighted with white ellipses, with their sizes enlarged by a factor of 1.5 relative to the measured values for clarity.}
\label{fig:sbackground}
\end{figure}

The inclusion of 7~m array improves not only the identification of compact sources, but also the recovery of their fluxes. We compared the integrated fluxes and peak intensities measured from source extraction performed on the different combined images. The integrated fluxes and peak intensities measured in the \texttt{12m7m} and \texttt{12m7mTP} images are broadly consistent, indicating that adding TP has a small effect on the measured fluxes of most compact sources. By contrast, measurements from the {\tt 12m}-only image are systematically lower, particularly for faint sources. Physically, this is consistent with the expectation that dense cores concentrate most of their mass on small spatial scales. 

Fig.~\ref{fig:sbackground} presents an example from the test field, illustrating how the smoothed background emission, $I_{\nu}^{\rm bg}$, and the source emission, $I_{\nu}^{\rm src}$, are successfully separated. The background component $I_{\nu}^{\rm bg}$ is essential for core mass estimates, as it enters as one term in the radiative transfer calculation. The background emission extracted from the \texttt{12m7m} image filters out structures larger than $\sim30\arcsec$, and therefore traces the dust emission distributed between dense cores. This component is denoted as intercore medium emission $I_{\nu}^{\rm ic}$. 

\section{Radiation transfer with background} \label{app:rt}

Considering three layers of medium, radiation transfer equation of dust thermal emission can be written as,
\begin{equation} \label{eq:fullRT}
    I^{\rm obs}_{\nu} = I^{\rm b}_{\nu} e^{-(\tau^{\rm c}_\nu+\tau^{\rm f}_\nu)} + B_{\nu}(T_{\rm c}) \left(1-e^{-\tau^{\rm c}_\nu}\right)e^{-\tau^{\rm f}_\nu} + I_\nu^{\rm f},
\end{equation}
where $I^{\rm b}_\nu$ is the intensity behind the core, $T_{\rm c}$ and $\tau^{\rm c}_\nu$ are the core dust temperature and optical depth, and $I^{\rm f}_\nu$ and $\tau^{\rm f}_\nu$ are emission and optical depth in front of the core. By definition, fore/background-related terms are spatially extended while core-related ones are compact. Eq.~\ref{eq:fullRT} can be transformed as
\begin{equation} \label{eq:sepRT}
    I^{\rm obs}_{\nu} =
    \underbrace{e^{-\tau_\nu^{\rm f}}\left(1-e^{-\tau_\nu^{\rm c}}\right)\left[B_\nu(T_{\rm c}) - I_\nu^{\rm b}\right]}_{\text{compact } I_\nu^{\rm com}}
    +
    \underbrace{I_\nu^{\rm b} e^{-\tau_\nu^{\rm f}}+ I_\nu^{\rm f}}_{\text{extended } I_\nu^{\rm ext}}
\end{equation}
where compact emission $I^{\rm com}_{\nu}$ and extended emission $I^{\rm ext}_{\nu}$ are separated. Such separation in formula aligns with what \texttt{getsf} isolates compact source emission $I^{\rm src}_\nu$ from extended emission $I^{\rm bg}_\nu$, as demonstrated in Fig.~\ref{fig:sbackground}. Here we should note that the terminology of ``background'' in \texttt{getsf} corresponds to the sum of the foreground and background emission in this section. We obtain two equations to solve:
\begin{subequations}\label{eq:sepRT_solve}
\begin{empheq}[left=\empheqlbrace]{align}
I^{\rm src}_\nu &= e^{-\tau_\nu^{\rm f}}\left(1-e^{-\tau_\nu^{\rm c}}\right)\left[B_\nu(T_{\rm c}) - I_\nu^{\rm b}\right]
\label{eq:solve1} \\
I^{\rm bg}_\nu &= I_\nu^{\rm b} e^{-\tau_\nu^{\rm f}} + I^{\rm f}_\nu
\label{eq:solve2}
\end{empheq}
\end{subequations}

Dense cores are assumed to be near the centers of their parent clouds. Under this assumption, the material in front of and behind a core is symmetric, so the dust optical depths are taken to be equal, $\tau_{\nu}^{\rm f}=\tau_{\nu}^{\rm b}$. The corresponding foreground and background intensities are therefore also assumed to be the same,
\begin{equation}
    I^{\rm f}_{\nu}=I^{\rm b}_{\nu}=B_\nu(T_{\rm env})\left(1-e^{-\tau_\nu^{\rm b}}\right).
\end{equation}
Solving Eq.~\ref{eq:solve2} gives
\begin{equation} \label{eq:tau_f}
    \tau_\nu^{\rm f} = \tau_\nu^{\rm b} = -\frac{1}{2}\ln\left(1-\frac{I_\nu^{\rm bg}}{B_\nu(T_{\rm env})}\right),
\end{equation}
where $T_{\rm env}$ is the environmental dust temperature. The total opacity summing up foreground and background is \begin{equation} \label{eq:tau_fb}
    \tau^{\rm fb}_\nu = 2\tau^{\rm b}_{\nu}
\end{equation}
Rewrite Eq.~\ref{eq:solve1} as expression to derive core opacity by
\begin{equation} \label{eq:tau_c}
    \tau_\nu^{\rm c} = - \ln\left\{ 1 - \frac{I^{\rm src}_\nu}{B_\nu(T_{\rm c}) - I^{\rm b}_\nu} \left[1-\frac{I_\nu^{\rm bg}}{B_\nu(T_{\rm env})}\right]^{-0.5} \right\}
\end{equation}
where the brightness behind the core is
\begin{equation} \label{eq:Ifb}
    I^{\rm b}_\nu = \frac{I^{\rm bg}_{\nu}}{\left[1-I^{\rm bg}_\nu/B_{\nu}(T_{\rm env})\right]^{0.5}+1}
\end{equation}

\begin{figure}[!h]
\centering
\includegraphics[width=\linewidth]{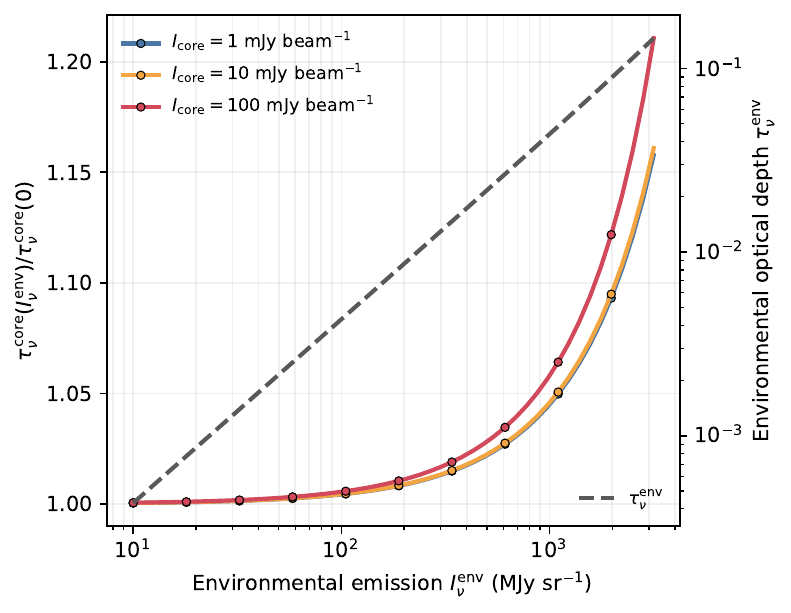}
\caption{{\bf Background emission effects on opacity estimates.} Gray curve shows total dust opacity summing up foreground and background of dense core. The color-coded curves show how much the opacity is underestimated when background emission is neglected, for core brightnesses of 1, 10, and 100~\mjybeam. }
\label{fig:backeffect}
\end{figure}

We calculated the dust opacity of dense cores (colored curves) as well as those in front of and behind dense cores (gray curve) in Fig.~\ref{fig:backeffect}. The dust temperatures of core environment and core are assumed to be the same, as a typical dust temperature of 20~K. For a background intensity below 1000~\mjysr\ and core peak brightnesses of 1--100~\mjybeam, omitting the foreground and background emission lowers the inferred core optical depth by less than approximately 6\% in this example. At a background intensity of 3000~\mjysr, the inferred optical depth including these terms is about 15--20\% higher than the zero-background estimate. This illustrates the growing importance of the background in brighter environments.

\section{Environmental kinematics structures} \label{app:sigma}

The environmental kinematic information is derived from the moment analysis of the C$^{18}$O (2-1) data cube obtained with APEX as part of the SEDIGISM survey \citep{Schuller2021}, which has a beam FWHM of 30\farcs2 and an rms noise of 1.0~K at a velocity resolution of 0.25~\kms. Prior to computing the moments, we apply a dilated masking approach (\url{https://github.com/tonywong94/maskmoment}) to suppress noise contributions, retaining only voxels above a specified signal-to-noise threshold in contiguous channels. We adopt the default parameters of the masking algorithm, except for a higher significance threshold of {\tt snr\_hi} = 9 to ensure robust emission identification. The moment maps are then calculated following standard definitions: the integrated intensity (moment 0) is obtained by summing the emission along the spectral axis, the intensity-weighted mean velocity (moment 1) corresponds to the first moment of the line profile, and the velocity dispersion (moment 2) is given by the square root of the intensity-weighted second central moment. This procedure ensures that the derived kinematic properties trace genuine emission structures while minimizing contamination from noise-dominated regions.

\begin{figure}[!h]
\centering
\includegraphics[width=\linewidth]{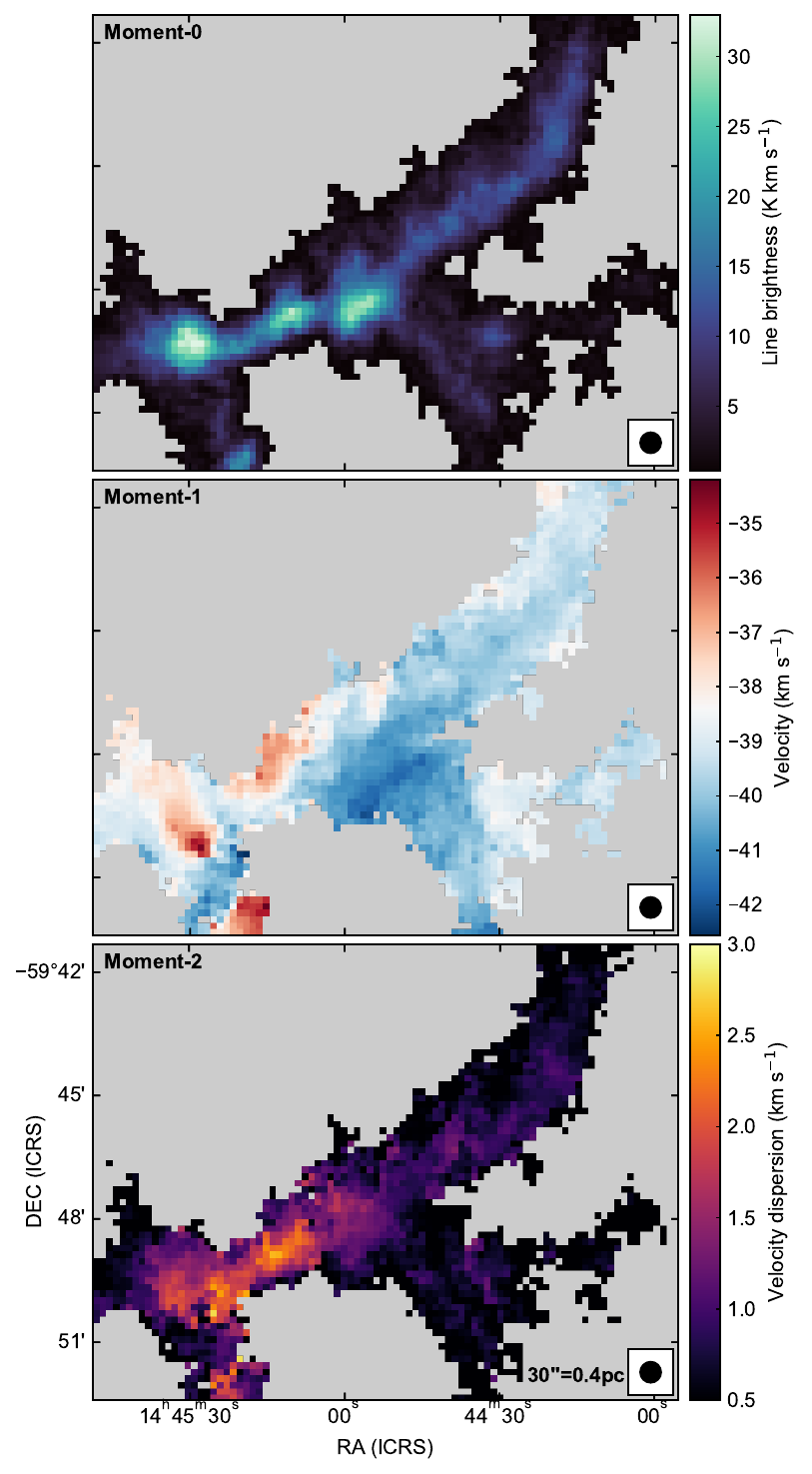}
\caption{{\bf Moment maps of the $\mathrm{C^{18}O}$ (2-1) emission.} The top, middle, and bottom panels show the integrated intensity (moment 0), intensity-weighted velocity field (moment 1), and velocity dispersion (moment 2), respectively. The beam FWHM is marked at the bottom right.}
\label{fig:c18o_moments}
\end{figure}

Our inspection identifies one coherent C$^{18}$O velocity component along the spine of G316.8. Multiple components along a sightline can contribute to the measured second moment, which we use as the observed line-of-sight velocity dispersion. The resulting moment maps are presented in Fig.~\ref{fig:c18o_moments}. At an angular resolution of $\sim30\arcsec$, or $\sim0.4$~pc, the global structure of the G316.8 filament is well resolved. The integrated intensity map reveals a prominent filamentary spine with a central hub. Along the spine, the velocity field exhibits a systematic redshift from the northern ($-40~\kms$) to the southern ($-37~\kms$) part, with a localized blueshift near the center. The velocity dispersion shows clear spatial variations, remaining close to $1~\kms$ in the north while increasing to $1$--$3~\kms$ toward the south. Previous studies \citep[e.g.,][]{Watkins2019} reported similar kinematic trends based on N$_2$H$^+$ (1--0), NH$_3$ (1,1), and $^{13}$CO (1--0) observations. Our measurements are consistent with these results, while providing improved angular resolution and more uniform sky coverage. We therefore adopt the SEDIGISM data as the basis for our analysis. 

\end{appendix}

\end{document}